\documentclass[aps,prl,twocolumn]{revtex4-2}
\usepackage{amsmath}
\usepackage{amssymb}
\usepackage{graphicx}
\usepackage{epstopdf}
\usepackage[colorlinks=true]{hyperref}

\usepackage{physics}
\usepackage{comment}

\newcommand{\sectionn}[1]{{\textit{#1}}}
\renewcommand\({\begin{equation}}	
\renewcommand\){\end{equation}}

\renewcommand\[{\begin{eqnarray}}	
\renewcommand\]{\end{eqnarray}}

\usepackage[dvipsnames]{xcolor}

\begin{document}

\title{Topological transport of deconfined hedgehogs in magnets}

\author{Ji Zou}
\author{Shu Zhang}
\email{suzy@physics.ucla.edu}
\author{Yaroslav Tserkovnyak}
\affiliation{Department of Physics and Astronomy, University of California, Los Angeles, California 90095, USA}

\begin{abstract}
We theoretically investigate the dynamics of magnetic hedgehogs, which are three-dimensional topological spin textures that exist in common magnets, focusing on their transport properties and connections to spintronics.  We show that fictitious magnetic monopoles carried by hedgehog textures obey a topological conservation law, based on which a hydrodynamic theory is developed.
We propose a nonlocal transport measurement in the disordered phase, where the conservation of the hedgehog flow results in a nonlocal signal decaying inversely proportional to the distance.
The bulk-edge correspondence between  hedgehog number and skyrmion number, the fictitious electric charges arising from magnetic dynamics, and the analogy between  bound states of hedgehogs in ordered phase and the quark confinement  in quantum chromodynamics are also discussed. Our study points to a practical potential in utilizing hedgehog flows for long-range neutral signal propagation or manipulation of skyrmion textures in three-dimensional magnetic materials.

\end{abstract}

\date{\today}
\maketitle

\underline{\sectionn{Introduction.}}|A main theme of spintronics is the utilization of spin degrees of freedom for information transmission and processing~\cite{Wolf1488, RevModPhys.76.323}, either using spin-polarized electric currents, or relying on spins alone to free the transport from Joule heating. 
Magnons, the quanta of  spin waves, have  been proposed to be promising data carriers in new computing technologies~\cite{RevModPhys.90.015005, Magnontransistor, Chumak:2015aa, Khitun_2010, Vogt:2014aa}. A detectable diffusive spin transport can be achieved via magnons in ordered magnetic insulators~\cite{Cornelissen:2015aa} or even spin-conserving fluctuations in paramagnets~\cite{Oyanagi:2019aa}. However, such spin currents typically decay  exponentially, once the propagation distance exceeds the spin-relaxation length~\cite{RevModPhys.76.323}. 
In alternative transport regimes, where signals are expected to decay algebraically, topology plays a crucial role~\cite{Yaroslavreview, TopologyinMagnetism, Ochoa:2019aa}.
Topological spin textures, such as chiral domain walls~\cite{sekwon2015domainwall}, vortices~\cite{PhysRevLett.108.167603, Chmiel:2018aa, jivortex, quantumvortex}, skyrmions~\cite{skyrmionreview, Jiang283, Hectorskyrmion}, hopfions~\cite{PhysRevLett.123.147203, Zangprl}, and hedgehogs~\cite{Fujishiro:2019aa, Tanigaki:2015aa, doi:10.1002/adma.201603227, PhysRevB.101.115144} are defined homotopically and are topologically protected~\cite{Yaroslavreview, TopologyinMagnetism, Ochoa:2019aa}.
Consequently, they are promising to sustain long-distance transport, even in the absence of local spin conservation. 


 \begin{figure}
  \includegraphics[scale=.23]{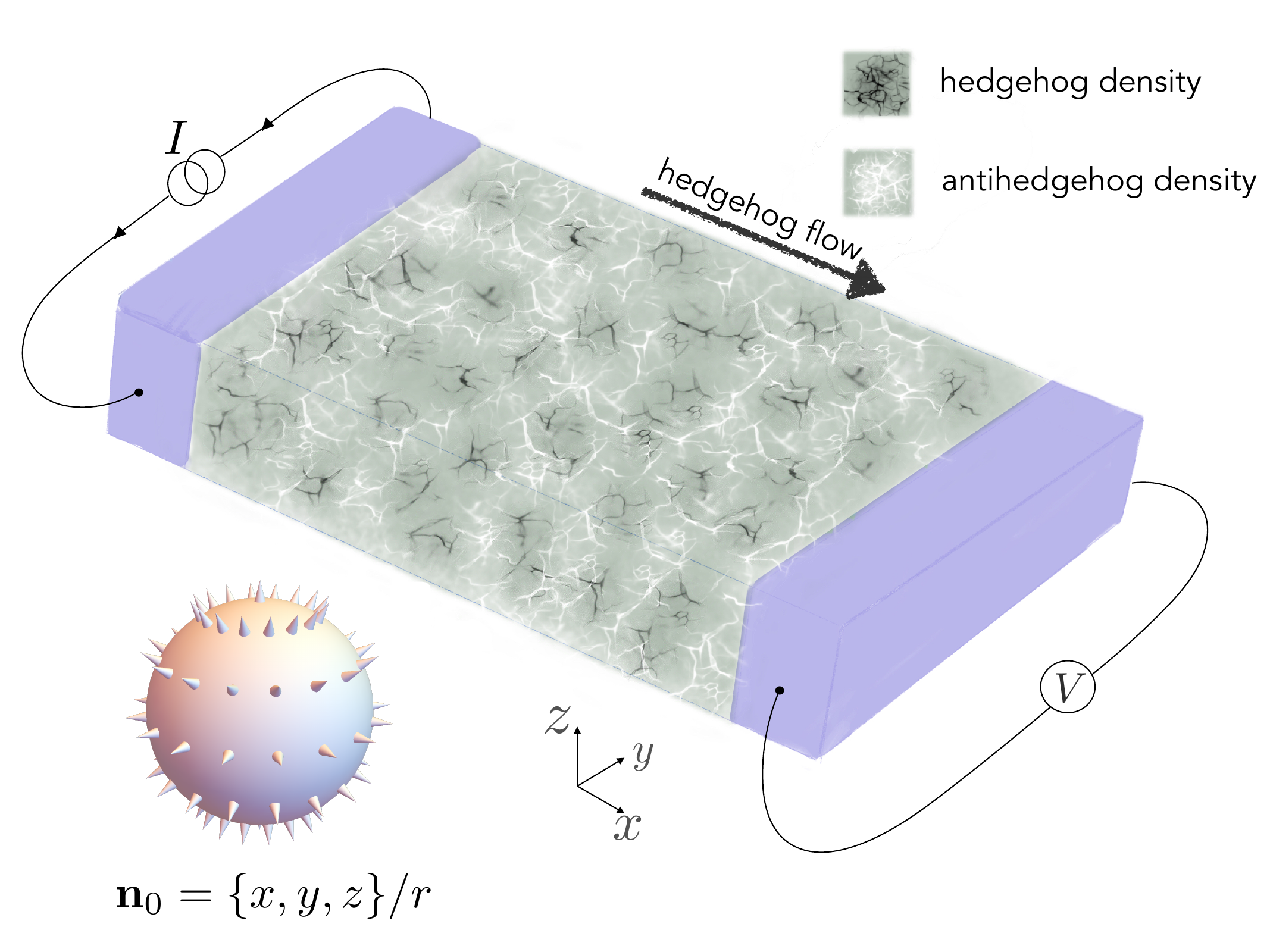}
   \caption{A schematic for nonlocal transport measurement of hedgehog currents in a three-dimensional insulating magnet. Two 
   metallic contacts are  bridged by  a magnetic insulator with hedgehog excitations. In the paramagnetic phase, hedgehogs are  free to diffuse, where black and white ripples stand respectively for delocalized hedgehog and antihedgehog densities. An applied electric current $I$ along $y$ within the left metal  transfers spin flow into the magnetic texture, which biases a hedgehog flow along $x$.  Reciprocally, the hedgehog flow reaching the right terminal builds up a detectable electric voltage $V$. The  nonlocal drag resistivity, $\varrho\propto V/I$, quantifies the efficiency of the topological hedgehog transport as well as their interfacial exchange coupling with conducting electrons. We also show a familiar example of a hedgehog $\vb{n}_0=\{x,y,z\}/r$.}
\label{fig1}
\end{figure}

While extensive studies have been devoted to spin textures in low dimensions, three-dimensional (3D) textures such as hedgehogs and hopfions are recently attracting more attention for their rich physics in topological phases~\cite{Fujishiro:2019aa, Tanigaki:2015aa, doi:10.1002/adma.201603227, PhysRevB.101.115144} and dynamic properties~\cite{PhysRevLett.123.147203, Zangprl, XXiao2016prb, Kanazawa:2016ab}. 
Hedgehogs exist inherently in 3D Heisenberg magnets. In contrast to 3D skyrmions~\cite{PhysRevB.100.054426} and hopfions~\cite{PhysRevLett.123.147203, Zangprl}, which can be annihilated by shrinking them down to the size of the atomic spacing without affecting spins far away, hedgehogs cannot be removed via local surgeries. The hedgehog flow can therefore be more stable against thermal fluctuations, and has potential applications in memory, logic devices, and energy storage~\cite{Allwood1688, Fert:2013aa, Parkin190, RevModPhys.80.1517,PhysRevLett.121.127701,energystorage}.


In this paper, we explore both topological and energetic properties of magnetic hedgehogs in 3D Heisenberg ferromagnets to investigate their long-distance transport, the viability of which is considered from the following three aspects. A topological conservation law, which is valid in both the magnetically ordered and disordered phase, defines the framework of a hydrodynamic description of hedgehog currents.
While (anti)hedgehogs are bound by a linear potential energy in the magnetically ordered phase, they become deconfined and hence mobile in the paramagnetic phase.
We propose a nonlocal transport measurement in the setup as shown in Fig.~\ref{fig1}.  A conserved hedgehog flow can be driven by a transverse electric current applied at an adjacent metal contact, resulting in a nonlocal signal decaying inversely proportional to the system length in the flow direction.


\underline{\sectionn{Topological conservation law.}}|Let us first consider  a 3D insulating ferromagnet without accounting for its detailed energetics, but focusing on   topological aspects of its vectorial order parameter $\vb{n}(\vec{r},t)$, where the bold face is used for axial vectors and the vector arrow marks polar vectors.  In the ordered phase, the collinear magnetic order can be described by the directions of $\vb{n}$  assuming $|\vb{n}| = 1$. This would render a sphere order-parameter space $S^2$, with a nontrivial second homotopy group $\pi_2(S^2)=\mathbb{Z}$~\cite{nkhr}. A  point defect, named hedgehog, with an integer-valued topological charge can correspondingly be identified in $\vb{n}(\vec{r},t)$. A familiar example for a hedgehog placed at the origin is  $\vb{n}_0=\{x,y,z\}/|\vec{r}|$. 

In the paramagnetic phase, the corresponding (coarse-grained) vector field $\vb{n}(\vec{r},t)\in \mathbb{R}^3$  realizes an $\mathbb{R}^3\rightarrow \mathbb{R}^3$ mapping at any given time $t$. This field texture is devoid of defects characterized by the aforementioned quantized charges, since the homotopy group $\pi_2(\mathbb{R}^3)$  is trivial \cite{nkhr}. Nevertheless, the smooth field $\vb{n}(\vec{r},t)$ exhibits a topological hydrodynamics governed by the \textit{topological conservation law} $\partial_\mu j^\mu=0$ (with the Einstein summation implied over the Greek indices: $\mu=0,1,2,3\leftrightarrow t,x,y,z$), where
\( j^\mu=\epsilon^{\mu\nu\alpha\beta}\partial_\nu\vb{n}\cdot (\partial_\alpha \vb{n}\times \partial_\beta\vb{n})/8\pi. \label{current} \)
Here, $\epsilon^{\mu\nu\alpha\beta}$ is the Levi-Civita symbol with convention $\epsilon^{0123}=1$. 
The conserved {(topological) charge} within a bulk $\Omega$ is
\( \mathcal{Q}\equiv\int_{\Omega}dxdydz\, j^0=\frac{1}{8\pi}\int_{\partial\Omega} dx^j\wedge dx^k\,\vb{n}\cdot (\partial_j\vb{n}\times \partial_k\vb{n}), \label{stokes} \)
which equals the skyrmion number at boundary $\partial\Omega$, according to the generalized Stokes' theorem \cite{nkhr}. 
We recognize that  charge $\mathcal{Q}$ is precisely the hedgehog number (thus $j^\mu$  is the hedgehog current) in the ordered phase, with the last equality in Eq.~(\ref{stokes}) defining the degree of the $S^2\rightarrow S^2$ mapping on the boundary. Our simple example $\vb{n}_0$ yields $j^0 \rightarrow \delta(\vec{r})$ and thus $\mathcal{Q}=1$. Here, we remark that the core should be regularized. There is no true singularity in our treatment.   In the paramagnetic phase, $\mathcal{Q}$ is no longer quantized due to  fluctuations in the magnitude of $\vb{n}$. Regardless, the hedgehog current~(\ref{current})  is conserved~\cite{sm}, which sets the stage for the topological hydrodynamics of hedgehogs at an arbitrary temperature. The conservation law also holds in the lattice limit, with proper discretized definitions~\cite{sm}. Hereafter, we refer to $j^\mu=(j^0,\vb{j})$ as hedgehog density (and flux), irrespective of the temperature.

We stress that, in contrast to two-dimensional skyrmions~\cite{skyrmionreview, Jiang283, Hectorskyrmion} or three-dimensional Shankar skyrmions~\cite{PhysRevB.100.054426}, which can be created and annihilated locally, the conservation law of hedgehogs is immune to local fluctuations and therefore applicable also in the paramagnetic phase~\cite{sm}. This robustness of hedgehog flow underpins the hedgehog hydrodynamics.  

Equation (\ref{stokes}) establishes a bulk-edge correspondence, indicating that the total hedgehog number in a bulk interior can fluctuate only by flowing in and out through its boundary. This, in turn, is associated with a corresponding change in the skyrmion number on the boundary, acting as a fingerprint of the hedgehog flow. A close analog in lower dimensions has been thoroughly studied in the context of superfluid phase slips, where the winding number associated with one-dimensional $XY$ textures can be changed by a transverse passage of planar vortices.
The 3D bulk-edge correspondence (\ref{stokes}) manifests when a skyrmion density unwinds or reversely builds up as a thread of a hedgehog current passes through, which has been verified experimentally~\cite{Milde1076}.  

\underline{\sectionn{Topological Maxwell equations.}}|We provide, in this section, another formulation of the conservation law as topological Maxwell equations, making connections to the well-known emergent electromagnetic fields associated with  generic spin textures~\cite{linearmomentum, Nagaosa_2012, PhysRevB.80.184411, TATARA2019208, Schulz:2012ab, PhysRevLett.122.187203, PhysRevB.87.024402}. 
The  divergence-free condition $\partial_\mu j^\mu=0$ can be automatically satisfied by defining the current $j^\mu$ as a curl of a rank-2 antisymmetric Maxwell field-strength  tensor
\( \mathcal{F}_{\alpha\beta}\equiv\vb{n}\cdot (\partial_\alpha \vb{n}\times \partial_\beta \vb{n})/4\pi, \label{field}   \)
whose components are the familiar electromagnetic fields:
\( E^i=\vb{n}\cdot (\partial_t\vb{n}\times \partial_i\vb{n})/4\pi, \,\,\,\, \epsilon_{ijk}B^k=\vb{n}\cdot (\partial_i\vb{n}\times \partial_j\vb{n})/4\pi.  \label{EB}  \)
The hedgehog current (\ref{current})  can therefore be recast into the form of the Maxwell equations: 
\( \epsilon^{\mu\nu\alpha\beta}\partial_\nu \mathcal{F}_{\alpha\beta}/2=j^\mu,  \,\, \partial_\mu \mathcal{F}^{\mu\nu}=j_e^\nu. \label{ME}\)
The second equation defines the electric 4-current, which is also conserved: $\partial_\mu j_e^\mu=0$,  following  from  the antisymmetric property of $\mathcal{F}$.  

Note that  fictitious electric and magnetic charges (as sources for $\vec{E}$ and $\vb{B}$)  have the same symmetries as the real electric and magnetic charges under both time-reversal and parity operations. The magnetic hedgehog with a quantized topological charge can be identified as a Dirac monopole.
In the ordered phase, the quantization can also be understood from the view of the U(1) gauge structure of magnons. For a Heisenberg magnet, one may regard a fixed spin texture $\vb{n}(\vec{r})$ to spontaneously break the SU(2) symmetry of the spin algebra. The excitations thus have a smooth part, which can be viewed as describing regular spin waves, and a (singular) topological part, such as hedgehogs in our case. 
Therefore, the magnetic charges are quantized on a U(1) kernel~\cite{aspectofsymmetry, linearmomentum, TATARA2019208},
reminiscent of the 't~Hooft and Polyakov approaches to the non-Abelian Higgs model~\cite{gaugefieldstring}. 
For the (fictitious) electric sector, both the electric field and the electric charge density emerge solely out of the dynamics of the field configuration.
According to the Gauss law, the total electric charge associated with a general local dynamics vanishes (see Supplemental Material~\cite{sm} for electric charge distributions resulted from global dynamics).
We restrict our discussion hereafter to the magnetic part.


\underline{\sectionn{Confinement and Deconfinement.}}|We now turn to the energetics and physical dynamics of hedgehogs in Heisenberg magnets. Hedgehogs are confined in the ordered phase, with the potential energy of a hedgehog-antihedgehog pair growing linearly with their separation~\cite{brezis1986}. One can imagine a string of tension $4\pi\mathcal{A}$  tying  them, where $\mathcal{A}$ is the exchange stiffness of the magnetic material [see Eq.~(\ref{potential}) below]. The confinement is also expected from evaluating the potential energy of a single hedgehog,
\( \mathcal{U}=
4\pi \mathcal{A} R, \label{potential} \)  where $R$ is the size of the system.  One can directly check this for $\vb{n}=\vb{n}_0$ placed at the origin of a sphere of radius $R$.  
For a hedgehog-antihedgehog pair, all flux from the hedgehog must end at the antihedgehog, forming a flux tube to minimize the energy. 
Such a system could realize an experimentally accessible analogy to quark confinement in QCD~\cite{RevModPhys.55.775,RevModPhys.49.267}.



 \begin{figure}
  \includegraphics[scale=.17]{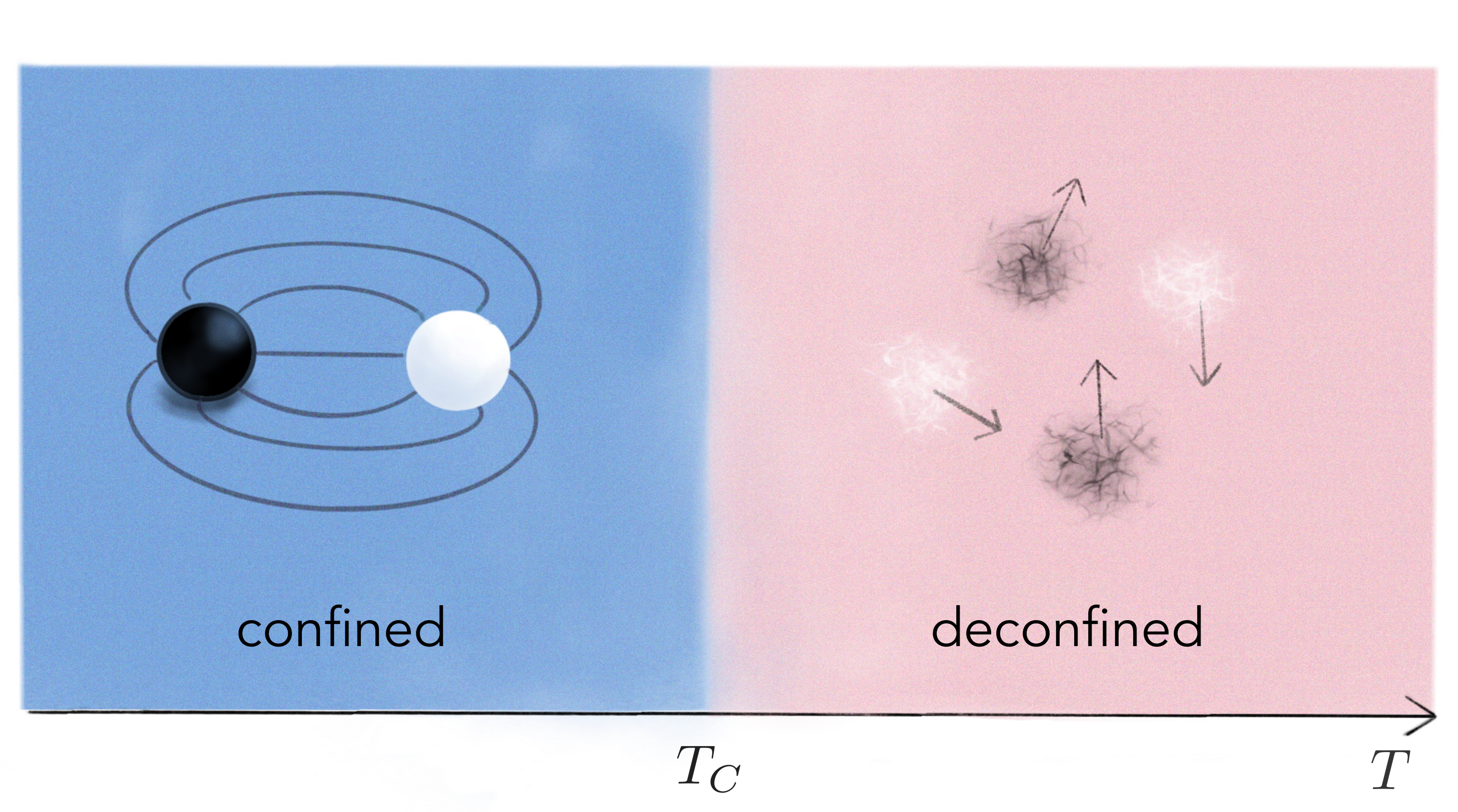}
   \caption{Different phases of hedgehogs. At temperatures above the Curie temperature $T_C$, hedgehogs (black ripples) and antihedgehogs (white ripples) carrying nonquantized topological charges proliferate and become mobile.  In the ordered phase, hedgehogs are confined by a linear potential analogous to the quark confinement in QCD. They are singular quantized objects, represented by black and white spheres.   } 
   \label{fig2}
\end{figure}


In the paramagnetic phase, hedgehogs deconfine naturally due to the absence of long-range  correlations.  Conceptually, the transition to the magnetically disordered phase can be thought of as a result of proliferation of hedgehogs,
which form a two-component hedgehog-antihedgehog plasma (while topological charges are no longer quantized) as illustrated in Figs.~\ref{fig1} and \ref{fig2}. This extends our analogy with the QCD picture: Quarks deconfine and form a so-called quark-gluon plasma at high temperatures where the chiral symmetry is restored and long-range correlation is melted away~\cite{RevModPhys.58.1021, quarksoup, PhysRevC.78.034919, PhysRevLett.94.172301}. The hedgehog system is therefore a promising alternate to study QCD theories in condensed matter systems. Previously considered was the magnetic monopole confinement in superconductors due to Meissner effect~\cite{Nielsen:1973cs, Nambuconfine, HOOFT1974276, PhysRevD.37.524, CARDOSO2008337}, where monopoles enter a deconfined phase as the temperature increases, accompanying the superconductor-insulator transition.

We remark that hedgehogs are not always energy-costly excitations as in Heisenberg magnets. $\text{MnSi}_{1-x}\text{Ge}_x$ has a stable phase of a hedgehog-antihedgehog lattice~\cite{Fujishiro:2019aa, Tanigaki:2015aa, doi:10.1002/adma.201603227}, where there is evidence that four- and six-spin interactions may play an important role~\cite{higherorder, PhysRevB.101.144416, PhysRevB.90.174432}. Here, we formally adopt a nonlocal term, which is particularly compatible with our fictitious electromagnetic formalism, into our Heisenberg Hamiltonian:
\( \mathcal{U}=\int d^3\vec{r}\,\, \Big[   \frac{\mathcal{A}}{2} (\vec{\nabla}\vb{n})^2+  \frac{\mathcal{C}}{2} (\partial_i\vb{n}\times \partial_j\vb{n})^2\Big],  \label{pe} \)
where Einstein summation is implied over the $i,j$ and $\mathcal{C}$ is a phenomenological parameter. Other quartic terms, such as $(\vec{\nabla} \vb{n})^4$, are also present in principle, but the above $\mathcal{C}$ term is of special interest to us. It resembles the Maxwell magnetic field energy $\propto \vb{B}^2$ (in the ordered phase, s.t. $|\vb{n}|\approx1$). The total potential energy for a hedgehog is thus $ \mathcal{U}\sim \mathcal{A}R- \mathcal{C}/R$, which indicates that hedgehogs are free at a small distance $r\ll \sqrt{\mathcal{C}/\mathcal{A}}$, where the Coulombic term dominates. This phenomena  mimics the asymptotic freedom in QCD \cite{RevModPhys.55.775,RevModPhys.49.267}. Likewise, in a system of small size, the hedgehog confinement becomes insignificant when $\sqrt{\mathcal{C}/\mathcal{A}}$ is comparable to $R$. 


\underline{\sectionn{Biased hedgehog flow.}}|At the heart of the hedgehog transport is the (desired) capability of driving and manipulating the hedgehog current.  
We focus on the dynamics near the Curie temperature of a ferromagnet, where the hedgehogs get deconfined, while the thermal fluctuations are still not too violent atomistically.
The spin dynamics can be described by
 $\partial_t \vb{n}=-\Gamma \vb{H}$, where we allow the magnitude $|\vb{n}|$ to fluctuate.  $\vb{H}=\delta F/\delta \vb{n}$ is the effective field (with $F$ being the free energy of the ferromagnet) and $\Gamma$ is a phenomenological dissipative coefficient.
As illustrated in  Fig.~\ref{fig1}, we apply a linear electric current density (i.e., current per $z$ thickness) $\vec{\mathcal{J}}=\mathcal{J}\hat{y}$ in the left metal contact. Given that the mirror-reflection symmetry is broken along the $z$ axis, 
  symmetry considerations~\cite{YSST,Hectorskyrmion,Sekwonvor, PhysRevB.100.054426} suggest that a spin flow can be transferred (per unit area) into the magnetic texture, in the form:
\( \vb*{\mathfrak{J}}=\frac{3\hbar }{8e\pi}    (\vec{\zeta} \cdot \vec{\nabla}\vb{n}) \times  (\vec{\mathcal{J}}\cdot \vec{\nabla} \vb{n})  ,  \label{torque} \)
where   $\vec{\zeta}=\zeta \hat{z}$ is a phenomenologically constructed vector (with the dimension of length), which reflects the mirror-symmetry breaking in the $z$ direction. Note that the spin transfer  $\vb*{\mathfrak{J}}$  is isotropic in spin space and thus does not rely microscopically on the presence of a spin-orbit coupling.  We also remark that $\vb*{\mathfrak{J}}$ is constructed phenomenologically based on symmetry considerations. Its microscopic origin, magnitude, and direct experimental signatures remain to be explored.

In the presence of this spin transfer, we have $\partial_t \vb{n}=-\Gamma \vb{H}+ \vb*{\mathfrak{J}}/s$, where $s$ is the saturated spin density. The rate of change of the free energy density is thus $P=\partial_t \vb{n}\cdot \vb{H}=\vb*{\mathfrak{J}} \cdot \partial_t \vb{n}/s\Gamma- (\partial_t \vb{n})^2/\Gamma$.
Here, the second term is the Rayleigh dissipation and the first term yields the total work done by the left contact upon magnetic dynamics:
\( W=\int dydzdt \,\, \frac{ \vb*{\mathfrak{J}} \cdot \partial_t \vb{n}}{s\Gamma}=\frac{  \hbar  \zeta \mathcal{J}  }{2es\Gamma} \mathcal{Q}.  \label{work}  \)
Note the work  is proportional to  the topological charge $\mathcal{Q}$. The setup in Fig.~\ref{fig1}  therefore discriminates between topological charges of opposite signs, biasing a net hedgehog current.
We also remark that  mirror symmetry in the $z$ direction is the minimal symmetry we need to break, in order for $\vec{\zeta}$  to effectively realize a polar vector in Eq.~(\ref{torque}).
 As a symmetry-allowed process, a hedgehog current can be generated naturally as a channel to release the energy associated with electron dynamics in the left contact (where an electric current is applied, see Fig.~\ref{fig1}). 
 The electric current in the (left) metal contact  provides an effective boundary bias for the hedgehog flow into the bulk,  effectively  establishing a local chemical potential for  hedgehogs:
\( \bar{\mu}_L\equiv \fdv{W}{\mathcal{Q}}=\frac{\hbar  \zeta \mathcal{J} }{2es\Gamma}.\)

\underline{\sectionn{Nonlocal spin drag.}}|We are now ready to study the transport  of hedgehogs in the geometry depicted in Fig.~\ref{fig1}. To this end, we operate the system in the paramagnetic phase, such that hedgehogs become mobile, rendering a finite effective  hedgehog conductivity $\sigma$.    We employ an Ohmic constitutive relation $j_x=-\sigma \partial_x \mu$ within the magnet. At the left boundary, using an electric current, the chemical potential for hedgehogs is raised by $\bar{\mu}_L$. At the right terminal, which serves as the ground, the natural chemical potential vanishes,  as hedgehogs can freely go in and out through the right boundary.  The hedgehog flow in $y$ and $z$ directions are nonvanishing in general~\cite{sm}. Here we  assume translational invariance along these two directions and focus on  dynamics in $x$ direction.
By invoking the reaction-rate theory \cite{reactionrate}, we obtain the hedgehog inflow and outflow at the left and right boundaries:
\( j^{L}_x=2\gamma_{L}(\bar{\mu}_{L}-\mu_{L})/k_BT,\,\,\, j^{R}_x=2\gamma_{R}\mu_{R}/k_BT,  \)
in  linear response. Here, $\mu_{L,R}\equiv \mu(x=0,L)$ is the chemical potential at two ends of the  magnet, where $L$ is its length  along the $x$ direction.  $\gamma_{L,R}\sim \nu_{L,R}e^{-E_0/k_BT}$ is the equilibrium injection rate of hedgehogs at the respective boundaries, in terms of the attempt frequencies $\nu_{L,R}$ and an effective energy barrier $E_0$ governed by the core energy.  Continuity of the hedgehog flux establishes a steady-state current
\( j_x= \frac{\hbar \mathcal{J}  \zeta /2es\Gamma}{L/\sigma+k_BT/\gamma}, \label{cur}   \)
where we took $\gamma_L=\gamma_R=\gamma$ for simplicity. $k_BT/\gamma$ is the boundary impedance. $L/\sigma$ is the bulk impedance, scaling linearly with the system size, which dominates in the thermodynamic limit $L\rightarrow \infty$.  This is similar to the ordinary Ohmic impedance for a conserved electric current.
  The hedgehog conductivity $\sigma=\rho_0D/k_BT$ is highly tunable via temperature~\cite{sm}, where $\rho_0$ and $D$ are the background hedgehog density and the diffusion constant, respectively.  In the paramagnetic phase ($T\gtrsim J/k_B$ with $J$ being the exchange energy), let us consider the limiting case where the order parameter varies on the atomic scale. Here, we estimate $\rho_0\sim 1/a^3$ and $D\sim Ja^2/\hbar$, with $a$ being the lattice spacing. This yields the optimal conductivity $\sigma\sim 1/a\hbar$ in the paramagnetic phase,  while   $\rho_0\propto e^{-E_0/k_BT}$ and thus $\sigma$      is exponentially small  in the ordered phase.

The hedgehog current reaching the right terminal exerts a pumping electromotive force on the metal contact~\cite{linearmomentum, PhysRevB.77.134407, PhysRevB.80.184411}, which  is determined by invoking the Onsager reciprocity~\cite{PhysRev.37.405}: $\vec{\varepsilon}=\hbar\, \vb{j}  \times\vec{\zeta} /2es\Gamma$. This leads to a finite nonlocal drag resistivity (whose sign is opposite to a viscous drag):
\( \varrho\equiv\frac{\varepsilon}{\mathcal{J}}= \frac{( \hbar  \zeta/2es\Gamma )^2}{L/\sigma+k_BT/\gamma}, \label{res}  \)
which is defined as the ratio of the detected voltage (per unit length) along $y$  at the right contact to the injected charge current density at the left contact. In the thermodynamic limit, the bulk impedance dominates and thus the  resistivity scales algebraically $\varrho\propto L^{-1}$.

In the ordered phase,  hedgehog currents vanish in linear response in the thermodynamic limit, while there can be  transport  in  finite-size magnets. The magnet behaves like a hedgehog dielectric, which can be polarized by the hedgehog chemical potential provided by the applied electric current in the  contact, like a  usual dielectric polarized by an electric field. This may be observed directly with the  coherent diffractive imaging~\cite{Donnelly:2017aa, Donnelly:2020aa} or by testing  (anti)skyrmion  structures on surfaces at  two ends~\cite{Romming636, doi:10.1002/adma.201500160, rougemaille2010magnetic}. 

Similar to the work on elastic response of skyrmion crystals~\cite{PhysRevB.96.020410}, one can study the transport feature due to the elasticity of a hedgehog lattice  across the hedgehog-lattice melting transition, in the geometry of Fig.~\ref{fig1}. To this end, the magnet is replaced by a hedgehog lattice, which should be experimentally accessible~\cite{Fujishiro:2019aa, Tanigaki:2015aa, doi:10.1002/adma.201603227}. 

Interesting physics emerges when one studies the interplay between electric currents and skyrmion dynamics \cite{Schulz:2012ab, StierPRL2017, Everschor_Sitte2017}, such as skyrmion Hall effect due to the  emergent magnetic flux associated with skyrmions. Similarly, we expect hedgehogs to  exhibit rich dynamics under electric currents, due to the possible complexity of emergent 3D  magnetic field associated with hedgehogs in general. 


\underline{\sectionn{Discussion.}}|We note that, though our discussion is based on insulating magnets, it applies equally well to conducting magnets, when the dimension along x is much larger than the other two in Fig.~\ref{fig1}, so that we can disregard an Ohmic drag between the side contacts.
Our symmetry arguments for the hedgehog injection can also be extended to antiferromagnetic systems, with the N{\'e}el order parameter, subject to a careful consideration of the magnetic space group.

There are two interesting open questions we did not address beyond  phenomenological level: the microscopic origin of the spin transfer (\ref{torque}) and a precise description of confinement-deconfinement transition of hedgehogs. We remark that the proposed nonlocal spin drag experiment in Fig.~\ref{fig1} also serves as a good testing platform of the deconfinement of hedgehogs in the paramagnetic phase. 

 The advances in  coherent diffractive imaging and vector-field tomography have proven promising for direct observations of 3D magnetic textures, as space and time resolutions are under improvement~\cite{Donnelly:2017aa, Donnelly:2020aa}. This  progress makes it possible for our theoretical concepts and proposals to be experimentally investigated for further understanding and future application of 3D topological spin textures.

Our study broadens the scope of 3D spintronics in at least two aspects: Hedgehogs, as commonly existing topological  textures in  magnets, are promising to support nonlocal transport in magnets and serve as information carriers. 
The study of the hedgehog current also offers us a handle to manipulate two-dimensional skyrmion textures, with various potential applications in skyrmionics. 
 Transverse skyrmion textures and their associated free energy can be loaded into a chiral magnet by biasing a hedgehog flux through it. If the induced skyrmion density can be preserved at lower temperatures, e.g. by the  Dzyaloshinskii-Moriya interaction, this would provide a 3D realization of the energy-storage proposal of Ref.~\cite{PhysRevLett.121.127701}. Being able to change the skyrmion number, the hedgehog flow can also be used to flip the polarity of a vortex core, while maintaining the vorticity.

\begin{acknowledgements}
We thank Se Kwon Kim, Julio Parra Martinez and Oleg Tchernyshyov for insightful discussions. This work is supported by NSF under Grant No. DMR-1742928.
\end{acknowledgements}

\onecolumngrid
\clearpage
\setcounter{equation}{0}
\renewcommand{\theequation}{S\arabic{equation}}
\appendix

{\centering
    \large{\textbf{{Supplemental Material for\\ ``Topological transport of deconfined hedgehogs in magnets"}}}
\par}

\bigskip
\bigskip
In this Supplemental Material, we present (i) proof of the topological conservation law, (ii) topological conservation law of hedgehogs on a lattice, (iii) electric charge distribution for global dynamics, and (iv) hedgehog conductivity from collective variable approach.
\subsection*{(i) Proof of the topological conservation law}
We prove the hedgehog current $j^\mu$ defined in the main text is conserved at arbitrary temperatures, which is the underpinning of the hedgehog hydrodynamics. For a general smooth texture $\vb{n}$, we can show,
\( \partial_\mu j^\mu= \epsilon^{\mu\nu\alpha\beta} \big[  \partial_\mu\partial_\nu \vb{n}\cdot (\partial_\alpha\vb{n}\times \partial_\beta\vb{n}) +\partial_\nu \vb{n}\cdot (  \partial_\mu\partial_\alpha\vb{n}\times \partial_\beta\vb{n})  +\partial_\nu \vb{n}\cdot (\partial_\alpha\vb{n}\times  \partial_\mu \partial_\beta\vb{n})    \big]/8\pi =0. \)
Here, all three terms vanish individually since $\epsilon^{\mu\nu}\partial_\mu\partial_\nu=\epsilon^{\mu\nu}\partial_\nu\partial_\mu=\epsilon^{\nu\mu}\partial_\mu\partial_\nu=-\epsilon^{\mu\nu}\partial_\mu\partial_\nu=0$, where we change the order of partial derivatives in the first equal sign and we exchange  dummy indices $\mu,\nu$ in the second equal sign.  We remark that  we did not use the condition $|\vb{n}|=1$, meaning that the conservation of hedgehog currents $\partial_\mu j^\mu=0$ holds at arbitrary temperatures, irrespective of the magnitude fluctuations of $\vb{n}$. This is in contrast to skyrmions and hopfions, whose conservation laws are only true when the magnitude fluctuations of $\vb{n}$ are neglectable.


\subsection*{(ii) Topological conservation law of hedgehogs on a  lattice}
\begin{figure}[t]
  \includegraphics[scale=.3]{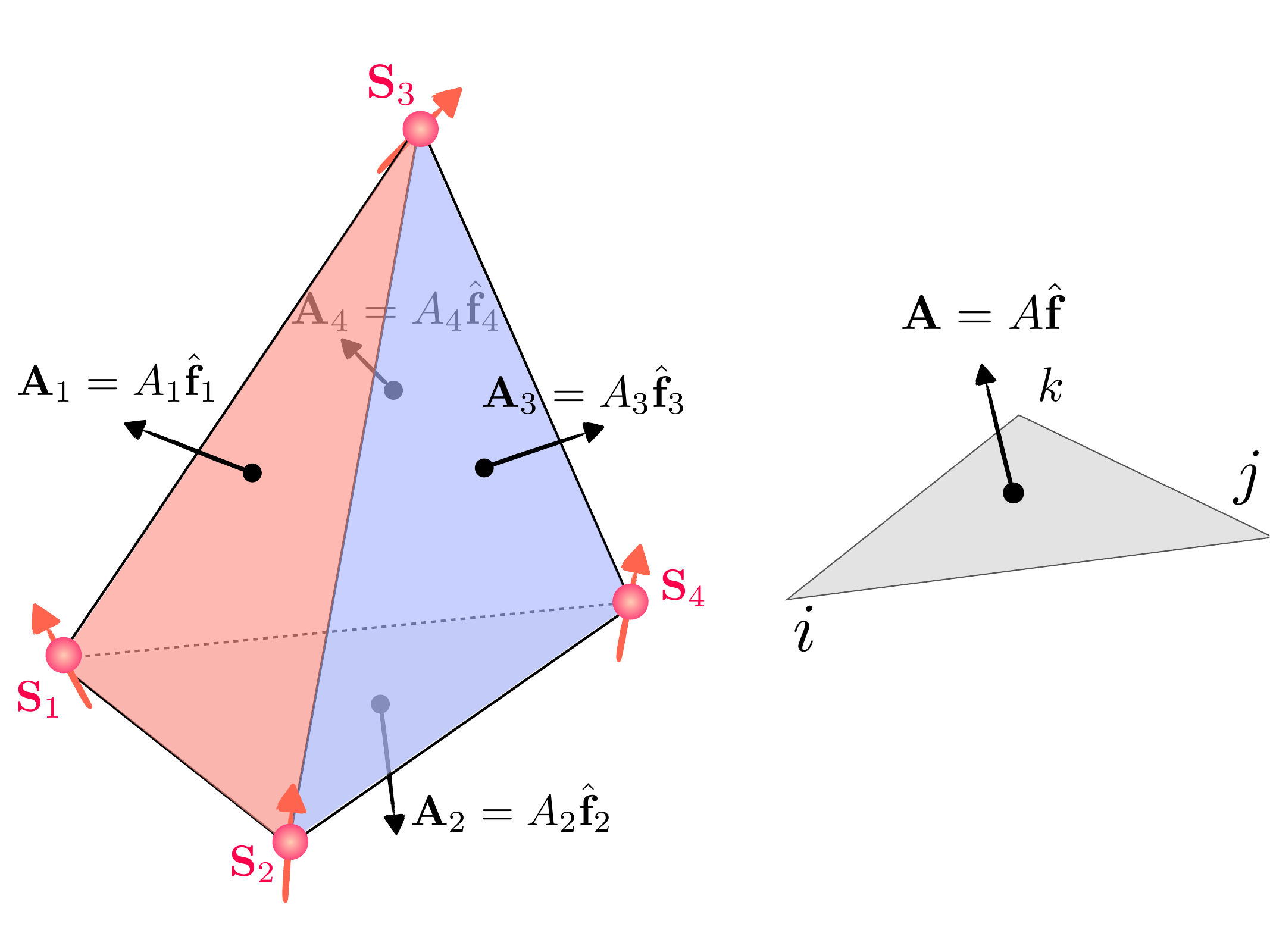}
   \caption{A tetrahedron severing as the  elementary building block of an arbitrary lattice. $\vb{S}_i$ is the spin operator at site $i$. A skyrmion number and a hedgehog flux \eqref{flux} can be defined for every facet $\vb{A}=A\hat{\vb{f}}$ with area $A$ and normal vector $\hat{\vb{f}}$. A hedgehog density \eqref{1} can be defined for every tetrahedron, where we choose the normal directions of all facets to be pointing outwards. } 
   \label{sm2}
\end{figure}

To construct a simple quantum theory, which reproduces the  classical hydrodynamics of hedgehogs in the classical limit of $\hbar\rightarrow 0$, let us consider a tetrahedron which  is   the elementary building block of any lattice (see Fig.~\ref{sm2}). Each site contains a  quantum spin  $\vb{S}=(S^x, S^y, S^z)$ of magnitude $S$ (in units of $\hbar$), obeying the standard SU(2) spin algebra $[S^a, S^b]=i \epsilon^{abc} S^c$ (spin operators sitting on different sites commute).

We first note that the  hedgehog 4-current  can be  recast as  $j^\mu=\epsilon^{\mu\nu\alpha\beta}\partial_\nu \mathcal{F}_{\alpha\beta}/2$ with $\mathcal{F}_{\alpha\beta} = \vb{n}\cdot (\partial_\alpha\vb{n}\times \partial_\beta\vb{n})/4\pi$. Its temporal  and spatial components are given by  $\rho=\nabla\cdot \vb{B}$ and $\vb{j}=-\partial_t \vb{B} +\nabla\times \vb{E}$ respectively, with the emergent magnetic field $B^i=\epsilon^{ijk}\vb{n}\cdot (\partial_j\vb{n}\times \partial_k \vb{n})/8\pi$ and the electric field $E^i=\vb{n}\cdot (\partial_t\vb{n} \times \partial_i \vb{n})/4\pi$. Every triangular  facet $\vb{A}=A\vb{f}$  ($A$ is the area and $\vb{f}$ is the normal vector) formed by sites $i, j, k$ (ordered in the right-hand fashion according to  $\vb{f}$ in Fig.~\ref{sm2}) can be associated  with  a skyrmion density and a hedgehog flux:
\(   B_{\vb{A}}  = \frac{c_{ijk}}{8\pi A}, \;\;\;\; \;\;\;\; j_{\vb{A}} =- \partial_t B_{\vb{A}} + \frac{\gamma_{ijk}}{A},   \label{flux} \)
where 
\(   c_{ijk}\equiv \frac{ \vb{S}_i\cdot \vb{S}_j\times \vb{S}_k}{S^3}, \;\;\;\; \;\;\;\;  \gamma_{ijk}\equiv   \vb{E}(\vb{r}_{ij}) \cdot \vb{r}_{ij} +\vb{E}(\vb{r}_{jk}) \cdot \vb{r}_{jk} +\vb{E}(\vb{r}_{ki}) \cdot \vb{r}_{ki},   \)
are the \textit{scalar chirality}  and the \textit{circulation} of $\vb{E}$ field along the facet. Here $\vb{r}_{ij}$ is the vector pointing from site $i$ to $j$ and $\vb{E}(\vb{r}_{ij})$ is the electric field on this link.
 By discretizing the electric field, we have \(  \vb{E}(\vb{r}_{ij}) \cdot \vb{r}_{ij} = \frac{\vb{S}_i\cdot \partial_t (\vb{S}_i +\vb{S}_j)\times \vb{S}_j  }{16\pi S^3} +\text{H.c.},   \)
where the time derivative should be understood to denote the Heisenberg commutator (for Hamiltonian $H$ and an arbitrary time-independent operator $\mathcal{O}$)
\( \partial_t \mathcal{O}\equiv \frac{i}{\hbar}[H, \mathcal{O}].   \)
We remark that the time derivative of spin does not commute with the bare spin operator, even on different sites, in general. 

By discretizing $\rho=\nabla\cdot \vb{B}$ in terms of the skyrmion densities (i.e., emergent $\vb{B}$ field) on the four facets, we can associate a hedgehog density 
\(  \rho= \frac{\sum_{i=1}^4 B_{\vb{A}_i} \, A_i}{V}= \frac{c_{123} + c_{142} +c_{243} +c_{134}}{8\pi V} , \label{1}  \)
to the tetrahedron in Fig.~\ref{sm2}, where $V$ is its volume.  Note definition~(\ref{1}) is proportional to the  total skyrmion number on all four facets (where we have chosen outwards normal vector as positive direction for defining the orientation). From this, we immediately conclude that the Stokes's theorem also holds at lattice level,
\( \sum_{\text{all cubes}} \rho V=\text{boundary skyrmions},  \label{stoke} \)
where all the inner facets cancel out and only boundary terms are left. Accordingly, for a fixed texture on the boundary, an arbitrary smooth field in the bulk yields the same net hedgehog number, irrespective of the details of the dynamics.

With these definitions of $\rho$ and $j_{\vb{A}}$, we can verify the hedgehog density and flux satisfy the continuity equation:
\( \partial_t ( \rho V) +\sum_{i=i}^4   j_{\vb{A}_i} \, A_i =0.     \)
One notes that there are two different contributions to the hedgehog flux \eqref{flux}. The first term, $- \partial_t B_{\vb{A}} $, is due to the change of the skyrmion density on the triangular  facet. The second term, $\gamma_{ijk}/A$, while  is inconsequential for the conservation law,  is important to ensure that $j_{\vb{A}}$ is a local physical current, which is consequential for energetic and Kubo considerations (when one attempts to establish chemical-potential bias of hedgehog and derive  its conductivity in linear response) \cite{quantumvortex}.

We emphasize that the conservation law for hedgehogs holds at quantum level (also in the classical limit), irrespective of the form of the Hamiltonian, which  indicates that it  is topological and is not rooted in any specific symmetry of the system. After coarse-graining process, we obtain $\vb{n}$ field out of $\vb{S}$ and hedgehog current $j^\mu$ in terms of $\vb{n}$. It is the conservation of hedgehog current at lattice level that determines the conservation of $j^\mu$ at arbitrary temperatures. 


\subsection*{(iii) Electric charge distribution for global dynamics}
For a translational motion, $\vb{n}(\vec{r},t)=\vb{n}_0(x,y,z-vt)$, which describes  a  hedgehog sliding along the $z$ axis with speed $v$, the fictitious electric field is 
\( \vec{E}=\frac{v(-y,x,0)}{4\pi r^3}. \)
This is precisely the Amp\`ere law for magnetic monopole, while the  electric charge density vanishes: $j^0_e=\nabla\cdot\vec{E}=0$. For a rotational motion, we take the example of $\vb{n}(\vec{r},t)=\vb{n}_0[\hat{\mathcal{R}}(\vb*{\omega} t)\vec{r}]$, which describes a hedgehog rotating with angular velocity $\vb*{\omega}=\omega\hat{z}$ (equivalent to precession in spin space),  where $\hat{\mathcal{R}}$ is the appropriate rotational matrix. This configuration yields a charge density with finite dipole moment proportional to the angular velocity:
\( j_e^0=-\frac{3\omega}{4\pi} \frac{z^3}{r^5}.  \)
The total electric charge remains zero in this example of an isotropic hedgehog configuration, but can be nonzero for a general rotating configuration, where the dynamic relation, $j^0_e\propto \omega$, remains. Therefore, if there is no overall rotation on thermal average, only the magnetic part $j^\mu$|hedgehog hydrodynamics|survives.

\subsection*{(iv) Hedgehog conductivity from collective variable approach}
Here we deduce the conductivity of hedgehogs, taking the view that they can be treated as  point-particles, where collective coordinate approach can be employed \cite{Stoneskyrmion, Hectorskyrmion}.  In the diffusion regime, the conductivity is $\hat{\sigma}=\rho_0\hat{\mu}=\rho_0\,\hat{D}/k_BT$, where $\hat{\mu}$ is the hedgehog mobility (related to diffusion constant $\hat{D}$ by the Einstein relation) and $\rho_0$ is the background hedgehog density \cite{1905AnP...322..549E}. The force acting on the hedgehog $\vec{F}$ is related with its velocity via the constitutive relation $\vec{v}=\hat{\mu}\cdot \vec{F}$. To work out the force due to the inherent spin Berry phase effect \cite{xiaogangwenqft}, we could look at the variation of the spin action:
\( \delta S= s \, \int d^3\vec{r} dt \,\, (\vb{n}\times \dot{\vb{n}})\cdot \delta\vb{n},  \)
 where $s= \hbar S/a^3$ is the saturated spin density coarse-grained from quantum-mechanical spin operators of length $S$ defined on a microscopic lattice of characteristic constant   $a$. By using the collective variables, 
 \( \vb{n}(\vec{r},t)=\vb{n}(\vec{r}-\vec{R}(t)),   \)
 one obtains
 \[  \delta S &=&  s \Big[ \int d^3\vec{r} dt \,\, \vb{n}\cdot (\partial_x\vb{n}\times \partial_y\vb{n}) \Big] \,\,\, (\dot{X}\delta Y -\dot{Y}\delta X)   \nonumber\\
                   &+&   s \Big[ \int d^3\vec{r} dt \,\, \vb{n}\cdot (\partial_y\vb{n}\times \partial_z\vb{n}) \Big] \,\,\, (\dot{Y}\delta Z -\dot{Z}\delta Y)   \nonumber\\ 
                   &+&  s \Big[ \int d^3\vec{r} dt \,\, \vb{n}\cdot (\partial_z\vb{n}\times \partial_x\vb{n}) \Big] \,\,\, (\dot{Z}\delta X -\dot{X}\delta Z),    \]
 from which we can write down the gyrotropic force acting on a hedgehog:
 \( \vec{F}= \vb*{\mathcal{B}} \times \vec{v}. \)
 Here, we introduce $ \vb*{\mathcal{B}} = (\mathcal{B}_x,\mathcal{B}_y, \mathcal{B}_z )$ with $\mathcal{B}_i\equiv 4\pi s \int d^3\vec{r}\, B_i$.
 
 To capture the damping effect, one need to start from Landau-Lifshitz-Gilbert equation and obtains the total force
 \( \vec{F}=   \vb*{\mathcal{B}}\times \vec{v} + \hat{\eta}\cdot \vec{ v}=\big[ ( \vb*{\mathcal{B}}\times) +  \hat{\eta}\cdot  \big]  \vec{v}. \label{force} \)
 Here,
 \( [\hat{\eta}]_{ij}=\alpha s \int d^3\vec{r}dt\,\, \partial_i\vb{n} \cdot \partial_j\vb{n}, \)
  is  a generalized viscosity tensor,  where $\alpha$ is the Gilbert damping.
 Therefore, the hedgehog conductivity can be formally written as
 \( \hat{\sigma} =\rho_0 \hat{\mu}=\rho_0\big[ ( \vb*{\mathcal{B}}\times) + \hat{\eta}\cdot  \big]^{-1}. \)
 One can write down the $\hat{\mu}$ (thus $\hat{\sigma}$) explicitly  by inverting the matrix equation (\ref{force}).     For example, the longitudinal conductivity in $x$ direction  is
\begin{footnotesize}
\( \sigma_{xx}= \frac{ \rho_0( \eta_{yy}\eta_{zz}-\eta^2_{yz}+ \mathcal{B}_x^2) }{  - \eta_{xx }\eta_{ yy}\eta_{ zz}-\eta_{ xx}\eta_{ yz}^2-\eta_{ yy}\eta_{ xz}^2-\eta_{ zz}\eta_{ xy}^2 +2\eta_{ xy}\eta_{ yz}\eta_{ zx}  +\eta_{xx}\mathcal{B}_x^2+\eta_{yy}\mathcal{B}_y^2 +\eta_{zz}\mathcal{B}_z^2 +2 \eta_{xy} \mathcal{B}_x\mathcal{B}_y+2 \eta_{xz} \mathcal{B}_x\mathcal{B}_z+2 \eta_{zy} \mathcal{B}_z\mathcal{B}_y         }  .   \)
\end{footnotesize} 
 
 Three important observations are in order. First, the resistivity tensor $\hat{\rho}=\hat{\sigma}^{-1}$  can be written as a sum of a symmetric tensor and an antisymmetric tensor. The former is  damping-related $\hat{\eta}$ which is dissipative while the latter is due to gyrotropic force (rooted in the spin Berry phase), which is nondissipative. Second, when the hedgehog is isotropic (e.g., $\vb{n}_0=\{x,y,z\}/r$), the gyrotropic force vanishes and $[\hat{\eta}]_{ij}=\eta \delta_{ij}$, where $\eta=4\pi s \int d^3\vec{r}dt\, \partial_x \vb{n}\cdot\partial_x \vb{n}$. Thus we have conductivity $\hat{\sigma}=\rho_0/\eta \delta_{ij}$.
  Third, our above derivation is based on collective coordinate approach. This approach breaks down at very high temperatures ($k_BT\geq J$ with $J$ being exchange energy), where we are not able to treat hedgehogs as point particles. In such a case, we expect the conductivity to be saturated and its temperature dependence to be weak. We approximate $\rho_0\sim 1/a^3$ and $D\sim Ja^2/\hbar$ based on basic dimension analysis. This would give us the optimal hedgehog conductivity $\sigma\sim 1/a\hbar$.


\begin{thebibliography}{75}%
\makeatletter
\providecommand \@ifxundefined [1]{%
 \@ifx{#1\undefined}
}%
\providecommand \@ifnum [1]{%
 \ifnum #1\expandafter \@firstoftwo
 \else \expandafter \@secondoftwo
 \fi
}%
\providecommand \@ifx [1]{%
 \ifx #1\expandafter \@firstoftwo
 \else \expandafter \@secondoftwo
 \fi
}%
\providecommand \natexlab [1]{#1}%
\providecommand \enquote  [1]{``#1''}%
\providecommand \bibnamefont  [1]{#1}%
\providecommand \bibfnamefont [1]{#1}%
\providecommand \citenamefont [1]{#1}%
\providecommand \href@noop [0]{\@secondoftwo}%
\providecommand \href [0]{\begingroup \@sanitize@url \@href}%
\providecommand \@href[1]{\@@startlink{#1}\@@href}%
\providecommand \@@href[1]{\endgroup#1\@@endlink}%
\providecommand \@sanitize@url [0]{\catcode `\\12\catcode `\$12\catcode
  `\&12\catcode `\#12\catcode `\^12\catcode `\_12\catcode `\%12\relax}%
\providecommand \@@startlink[1]{}%
\providecommand \@@endlink[0]{}%
\providecommand \url  [0]{\begingroup\@sanitize@url \@url }%
\providecommand \@url [1]{\endgroup\@href {#1}{\urlprefix }}%
\providecommand \urlprefix  [0]{URL }%
\providecommand \Eprint [0]{\href }%
\providecommand \doibase [0]{http://dx.doi.org/}%
\providecommand \selectlanguage [0]{\@gobble}%
\providecommand \bibinfo  [0]{\@secondoftwo}%
\providecommand \bibfield  [0]{\@secondoftwo}%
\providecommand \translation [1]{[#1]}%
\providecommand \BibitemOpen [0]{}%
\providecommand \bibitemStop [0]{}%
\providecommand \bibitemNoStop [0]{.\EOS\space}%
\providecommand \EOS [0]{\spacefactor3000\relax}%
\providecommand \BibitemShut  [1]{\csname bibitem#1\endcsname}%
\let\auto@bib@innerbib\@empty
\bibitem [{\citenamefont {Wolf}\ \emph {et~al.}(2001)\citenamefont {Wolf},
  \citenamefont {Awschalom}, \citenamefont {Buhrman}, \citenamefont {Daughton},
  \citenamefont {von Moln{\'a}r}, \citenamefont {Roukes}, \citenamefont
  {Chtchelkanova},\ and\ \citenamefont {Treger}}]{Wolf1488}%
  \BibitemOpen
  \bibfield  {author} {\bibinfo {author} {\bibfnamefont {S.~A.}\ \bibnamefont
  {Wolf}}, \bibinfo {author} {\bibfnamefont {D.~D.}\ \bibnamefont {Awschalom}},
  \bibinfo {author} {\bibfnamefont {R.~A.}\ \bibnamefont {Buhrman}}, \bibinfo
  {author} {\bibfnamefont {J.~M.}\ \bibnamefont {Daughton}}, \bibinfo {author}
  {\bibfnamefont {S.}~\bibnamefont {von Moln{\'a}r}}, \bibinfo {author}
  {\bibfnamefont {M.~L.}\ \bibnamefont {Roukes}}, \bibinfo {author}
  {\bibfnamefont {A.~Y.}\ \bibnamefont {Chtchelkanova}}, \ and\ \bibinfo
  {author} {\bibfnamefont {D.~M.}\ \bibnamefont {Treger}},\ }\href@noop {}
  {\bibfield  {journal} {\bibinfo  {journal} {Science}\ }\textbf {\bibinfo
  {volume} {294}},\ \bibinfo {pages} {1488} (\bibinfo {year}
  {2001})}\BibitemShut {NoStop}%
\bibitem [{\citenamefont {\ifmmode \check{Z}\else
  \v{Z}\fi{}uti\ifmmode~\acute{c}\else \'{c}\fi{}}\ \emph
  {et~al.}(2004)\citenamefont {\ifmmode \check{Z}\else
  \v{Z}\fi{}uti\ifmmode~\acute{c}\else \'{c}\fi{}}, \citenamefont {Fabian},\
  and\ \citenamefont {Das~Sarma}}]{RevModPhys.76.323}%
  \BibitemOpen
  \bibfield  {author} {\bibinfo {author} {\bibfnamefont {I.}~\bibnamefont
  {\ifmmode \check{Z}\else \v{Z}\fi{}uti\ifmmode~\acute{c}\else \'{c}\fi{}}},
  \bibinfo {author} {\bibfnamefont {J.}~\bibnamefont {Fabian}}, \ and\ \bibinfo
  {author} {\bibfnamefont {S.}~\bibnamefont {Das~Sarma}},\ }\href@noop {}
  {\bibfield  {journal} {\bibinfo  {journal} {Rev. Mod. Phys.}\ }\textbf
  {\bibinfo {volume} {76}},\ \bibinfo {pages} {323} (\bibinfo {year}
  {2004})}\BibitemShut {NoStop}%
\bibitem [{\citenamefont {Baltz}\ \emph {et~al.}(2018)\citenamefont {Baltz},
  \citenamefont {Manchon}, \citenamefont {Tsoi}, \citenamefont {Moriyama},
  \citenamefont {Ono},\ and\ \citenamefont
  {Tserkovnyak}}]{RevModPhys.90.015005}%
  \BibitemOpen
  \bibfield  {author} {\bibinfo {author} {\bibfnamefont {V.}~\bibnamefont
  {Baltz}}, \bibinfo {author} {\bibfnamefont {A.}~\bibnamefont {Manchon}},
  \bibinfo {author} {\bibfnamefont {M.}~\bibnamefont {Tsoi}}, \bibinfo {author}
  {\bibfnamefont {T.}~\bibnamefont {Moriyama}}, \bibinfo {author}
  {\bibfnamefont {T.}~\bibnamefont {Ono}}, \ and\ \bibinfo {author}
  {\bibfnamefont {Y.}~\bibnamefont {Tserkovnyak}},\ }\href@noop {} {\bibfield
  {journal} {\bibinfo  {journal} {Rev. Mod. Phys.}\ }\textbf {\bibinfo {volume}
  {90}},\ \bibinfo {pages} {015005} (\bibinfo {year} {2018})}\BibitemShut
  {NoStop}%
\bibitem [{\citenamefont {Chumak}\ \emph {et~al.}(2014)\citenamefont {Chumak},
  \citenamefont {Serga},\ and\ \citenamefont {Hillebrands}}]{Magnontransistor}%
  \BibitemOpen
  \bibfield  {author} {\bibinfo {author} {\bibfnamefont {A.~V.}\ \bibnamefont
  {Chumak}}, \bibinfo {author} {\bibfnamefont {A.~A.}\ \bibnamefont {Serga}}, \
  and\ \bibinfo {author} {\bibfnamefont {B.}~\bibnamefont {Hillebrands}},\
  }\href@noop {} {\bibfield  {journal} {\bibinfo  {journal} {Nature
  Communications}\ }\textbf {\bibinfo {volume} {5}},\ \bibinfo {pages} {4700}
  (\bibinfo {year} {2014})}\BibitemShut {NoStop}%
\bibitem [{\citenamefont {Chumak}\ \emph {et~al.}(2015)\citenamefont {Chumak},
  \citenamefont {Vasyuchka}, \citenamefont {Serga},\ and\ \citenamefont
  {Hillebrands}}]{Chumak:2015aa}%
  \BibitemOpen
  \bibfield  {author} {\bibinfo {author} {\bibfnamefont {A.~V.}\ \bibnamefont
  {Chumak}}, \bibinfo {author} {\bibfnamefont {V.~I.}\ \bibnamefont
  {Vasyuchka}}, \bibinfo {author} {\bibfnamefont {A.~A.}\ \bibnamefont
  {Serga}}, \ and\ \bibinfo {author} {\bibfnamefont {B.}~\bibnamefont
  {Hillebrands}},\ }\href@noop {} {\bibfield  {journal} {\bibinfo  {journal}
  {Nature Physics}\ }\textbf {\bibinfo {volume} {11}},\ \bibinfo {pages} {453}
  (\bibinfo {year} {2015})}\BibitemShut {NoStop}%
\bibitem [{\citenamefont {Khitun}\ \emph {et~al.}(2010)\citenamefont {Khitun},
  \citenamefont {Bao},\ and\ \citenamefont {Wang}}]{Khitun_2010}%
  \BibitemOpen
  \bibfield  {author} {\bibinfo {author} {\bibfnamefont {A.}~\bibnamefont
  {Khitun}}, \bibinfo {author} {\bibfnamefont {M.}~\bibnamefont {Bao}}, \ and\
  \bibinfo {author} {\bibfnamefont {K.~L.}\ \bibnamefont {Wang}},\ }\href@noop
  {} {\bibfield  {journal} {\bibinfo  {journal} {Journal of Physics D: Applied
  Physics}\ }\textbf {\bibinfo {volume} {43}},\ \bibinfo {pages} {264005}
  (\bibinfo {year} {2010})}\BibitemShut {NoStop}%
\bibitem [{\citenamefont {Vogt}\ \emph {et~al.}(2014)\citenamefont {Vogt},
  \citenamefont {Fradin}, \citenamefont {Pearson}, \citenamefont {Sebastian},
  \citenamefont {Bader}, \citenamefont {Hillebrands}, \citenamefont
  {Hoffmann},\ and\ \citenamefont {Schultheiss}}]{Vogt:2014aa}%
  \BibitemOpen
  \bibfield  {author} {\bibinfo {author} {\bibfnamefont {K.}~\bibnamefont
  {Vogt}}, \bibinfo {author} {\bibfnamefont {F.~Y.}\ \bibnamefont {Fradin}},
  \bibinfo {author} {\bibfnamefont {J.~E.}\ \bibnamefont {Pearson}}, \bibinfo
  {author} {\bibfnamefont {T.}~\bibnamefont {Sebastian}}, \bibinfo {author}
  {\bibfnamefont {S.~D.}\ \bibnamefont {Bader}}, \bibinfo {author}
  {\bibfnamefont {B.}~\bibnamefont {Hillebrands}}, \bibinfo {author}
  {\bibfnamefont {A.}~\bibnamefont {Hoffmann}}, \ and\ \bibinfo {author}
  {\bibfnamefont {H.}~\bibnamefont {Schultheiss}},\ }\href@noop {} {\bibfield
  {journal} {\bibinfo  {journal} {Nature Communications}\ }\textbf {\bibinfo
  {volume} {5}},\ \bibinfo {pages} {3727} (\bibinfo {year} {2014})}\BibitemShut
  {NoStop}%
\bibitem [{\citenamefont {Cornelissen}\ \emph {et~al.}(2015)\citenamefont
  {Cornelissen}, \citenamefont {Liu}, \citenamefont {Duine}, \citenamefont
  {Youssef},\ and\ \citenamefont {van Wees}}]{Cornelissen:2015aa}%
  \BibitemOpen
  \bibfield  {author} {\bibinfo {author} {\bibfnamefont {L.~J.}\ \bibnamefont
  {Cornelissen}}, \bibinfo {author} {\bibfnamefont {J.}~\bibnamefont {Liu}},
  \bibinfo {author} {\bibfnamefont {R.~A.}\ \bibnamefont {Duine}}, \bibinfo
  {author} {\bibfnamefont {J.~B.}\ \bibnamefont {Youssef}}, \ and\ \bibinfo
  {author} {\bibfnamefont {B.~J.}\ \bibnamefont {van Wees}},\ }\href@noop {}
  {\bibfield  {journal} {\bibinfo  {journal} {Nature Physics}\ }\textbf
  {\bibinfo {volume} {11}},\ \bibinfo {pages} {1022} (\bibinfo {year}
  {2015})}\BibitemShut {NoStop}%
\bibitem [{\citenamefont {Oyanagi}\ \emph {et~al.}(2019)\citenamefont
  {Oyanagi}, \citenamefont {Takahashi}, \citenamefont {Cornelissen},
  \citenamefont {Shan}, \citenamefont {Daimon}, \citenamefont {Kikkawa},
  \citenamefont {Bauer}, \citenamefont {van Wees},\ and\ \citenamefont
  {Saitoh}}]{Oyanagi:2019aa}%
  \BibitemOpen
  \bibfield  {author} {\bibinfo {author} {\bibfnamefont {K.}~\bibnamefont
  {Oyanagi}}, \bibinfo {author} {\bibfnamefont {S.}~\bibnamefont {Takahashi}},
  \bibinfo {author} {\bibfnamefont {L.~J.}\ \bibnamefont {Cornelissen}},
  \bibinfo {author} {\bibfnamefont {J.}~\bibnamefont {Shan}}, \bibinfo {author}
  {\bibfnamefont {S.}~\bibnamefont {Daimon}}, \bibinfo {author} {\bibfnamefont
  {T.}~\bibnamefont {Kikkawa}}, \bibinfo {author} {\bibfnamefont {G.~E.~W.}\
  \bibnamefont {Bauer}}, \bibinfo {author} {\bibfnamefont {B.~J.}\ \bibnamefont
  {van Wees}}, \ and\ \bibinfo {author} {\bibfnamefont {E.}~\bibnamefont
  {Saitoh}},\ }\href@noop {} {\bibfield  {journal} {\bibinfo  {journal} {Nature
  Communications}\ }\textbf {\bibinfo {volume} {10}},\ \bibinfo {pages} {4740}
  (\bibinfo {year} {2019})}\BibitemShut {NoStop}%
\bibitem [{\citenamefont {Tserkovnyak}(2018)}]{Yaroslavreview}%
  \BibitemOpen
  \bibfield  {author} {\bibinfo {author} {\bibfnamefont {Y.}~\bibnamefont
  {Tserkovnyak}},\ }\href@noop {} {\bibfield  {journal} {\bibinfo  {journal}
  {Journal of Applied Physics}\ }\textbf {\bibinfo {volume} {124}},\ \bibinfo
  {pages} {190901} (\bibinfo {year} {2018})}\BibitemShut {NoStop}%
\bibitem [{\citenamefont {Zang}\ \emph {et~al.}(2018)\citenamefont {Zang},
  \citenamefont {Cros},\ and\ \citenamefont {Hoffmann}}]{TopologyinMagnetism}%
  \BibitemOpen
  \bibinfo {editor} {\bibfnamefont {J.}~\bibnamefont {Zang}}, \bibinfo {editor}
  {\bibfnamefont {V.}~\bibnamefont {Cros}}, \ and\ \bibinfo {editor}
  {\bibfnamefont {A.}~\bibnamefont {Hoffmann}},\ eds.,\ \href@noop {} {\emph
  {\bibinfo {title} {Topology in Magnetism}}}\ (\bibinfo  {publisher} {Springer
  International Publishing},\ \bibinfo {year} {2018})\BibitemShut {NoStop}%
\bibitem [{\citenamefont {Ochoa}\ and\ \citenamefont
  {Tserkovnyak}(2019)}]{Ochoa:2019aa}%
  \BibitemOpen
  \bibfield  {author} {\bibinfo {author} {\bibfnamefont {H.}~\bibnamefont
  {Ochoa}}\ and\ \bibinfo {author} {\bibfnamefont {Y.}~\bibnamefont
  {Tserkovnyak}},\ }\href@noop {} {\bibfield  {journal} {\bibinfo  {journal}
  {Int. J. Mod. Phys. B}\ }\textbf {\bibinfo {volume} {33}},\ \bibinfo {pages}
  {1930005} (\bibinfo {year} {2019})}\BibitemShut {NoStop}%
\bibitem [{\citenamefont {Kim}\ \emph {et~al.}(2015)\citenamefont {Kim},
  \citenamefont {Takei},\ and\ \citenamefont
  {Tserkovnyak}}]{sekwon2015domainwall}%
  \BibitemOpen
  \bibfield  {author} {\bibinfo {author} {\bibfnamefont {S.~K.}\ \bibnamefont
  {Kim}}, \bibinfo {author} {\bibfnamefont {S.}~\bibnamefont {Takei}}, \ and\
  \bibinfo {author} {\bibfnamefont {Y.}~\bibnamefont {Tserkovnyak}},\
  }\href@noop {} {\bibfield  {journal} {\bibinfo  {journal} {Phys. Rev. B}\
  }\textbf {\bibinfo {volume} {92}},\ \bibinfo {pages} {220409} (\bibinfo
  {year} {2015})}\BibitemShut {NoStop}%
\bibitem [{\citenamefont {Chae}\ \emph {et~al.}(2012)\citenamefont {Chae},
  \citenamefont {Lee}, \citenamefont {Horibe}, \citenamefont {Tanimura},
  \citenamefont {Mori}, \citenamefont {Gao}, \citenamefont {Carr},\ and\
  \citenamefont {Cheong}}]{PhysRevLett.108.167603}%
  \BibitemOpen
  \bibfield  {author} {\bibinfo {author} {\bibfnamefont {S.~C.}\ \bibnamefont
  {Chae}}, \bibinfo {author} {\bibfnamefont {N.}~\bibnamefont {Lee}}, \bibinfo
  {author} {\bibfnamefont {Y.}~\bibnamefont {Horibe}}, \bibinfo {author}
  {\bibfnamefont {M.}~\bibnamefont {Tanimura}}, \bibinfo {author}
  {\bibfnamefont {S.}~\bibnamefont {Mori}}, \bibinfo {author} {\bibfnamefont
  {B.}~\bibnamefont {Gao}}, \bibinfo {author} {\bibfnamefont {S.}~\bibnamefont
  {Carr}}, \ and\ \bibinfo {author} {\bibfnamefont {S.-W.}\ \bibnamefont
  {Cheong}},\ }\href@noop {} {\bibfield  {journal} {\bibinfo  {journal} {Phys.
  Rev. Lett.}\ }\textbf {\bibinfo {volume} {108}},\ \bibinfo {pages} {167603}
  (\bibinfo {year} {2012})}\BibitemShut {NoStop}%
\bibitem [{\citenamefont {Chmiel}\ \emph {et~al.}(2018)\citenamefont {Chmiel},
  \citenamefont {Waterfield~Price}, \citenamefont {Johnson}, \citenamefont
  {Lamirand}, \citenamefont {Schad}, \citenamefont {van~der Laan},
  \citenamefont {Harris}, \citenamefont {Irwin}, \citenamefont {Rzchowski},
  \citenamefont {Eom},\ and\ \citenamefont {Radaelli}}]{Chmiel:2018aa}%
  \BibitemOpen
  \bibfield  {author} {\bibinfo {author} {\bibfnamefont {F.~P.}\ \bibnamefont
  {Chmiel}}, \bibinfo {author} {\bibfnamefont {N.}~\bibnamefont
  {Waterfield~Price}}, \bibinfo {author} {\bibfnamefont {R.~D.}\ \bibnamefont
  {Johnson}}, \bibinfo {author} {\bibfnamefont {A.~D.}\ \bibnamefont
  {Lamirand}}, \bibinfo {author} {\bibfnamefont {J.}~\bibnamefont {Schad}},
  \bibinfo {author} {\bibfnamefont {G.}~\bibnamefont {van~der Laan}}, \bibinfo
  {author} {\bibfnamefont {D.~T.}\ \bibnamefont {Harris}}, \bibinfo {author}
  {\bibfnamefont {J.}~\bibnamefont {Irwin}}, \bibinfo {author} {\bibfnamefont
  {M.~S.}\ \bibnamefont {Rzchowski}}, \bibinfo {author} {\bibfnamefont {C.~B.}\
  \bibnamefont {Eom}}, \ and\ \bibinfo {author} {\bibfnamefont {P.~G.}\
  \bibnamefont {Radaelli}},\ }\href@noop {} {\bibfield  {journal} {\bibinfo
  {journal} {Nature Materials}\ }\textbf {\bibinfo {volume} {17}},\ \bibinfo
  {pages} {581} (\bibinfo {year} {2018})}\BibitemShut {NoStop}%
\bibitem [{\citenamefont {Zou}\ \emph {et~al.}(2019)\citenamefont {Zou},
  \citenamefont {Kim},\ and\ \citenamefont {Tserkovnyak}}]{jivortex}%
  \BibitemOpen
  \bibfield  {author} {\bibinfo {author} {\bibfnamefont {J.}~\bibnamefont
  {Zou}}, \bibinfo {author} {\bibfnamefont {S.~K.}\ \bibnamefont {Kim}}, \ and\
  \bibinfo {author} {\bibfnamefont {Y.}~\bibnamefont {Tserkovnyak}},\
  }\href@noop {} {\bibfield  {journal} {\bibinfo  {journal} {Phys. Rev. B}\
  }\textbf {\bibinfo {volume} {99}},\ \bibinfo {pages} {180402} (\bibinfo
  {year} {2019})}\BibitemShut {NoStop}%
\bibitem [{\citenamefont {Tserkovnyak}\ and\ \citenamefont
  {Zou}(2019)}]{quantumvortex}%
  \BibitemOpen
  \bibfield  {author} {\bibinfo {author} {\bibfnamefont {Y.}~\bibnamefont
  {Tserkovnyak}}\ and\ \bibinfo {author} {\bibfnamefont {J.}~\bibnamefont
  {Zou}},\ }\href@noop {} {\bibfield  {journal} {\bibinfo  {journal} {Phys.
  Rev. Research}\ }\textbf {\bibinfo {volume} {1}},\ \bibinfo {pages} {033071}
  (\bibinfo {year} {2019})}\BibitemShut {NoStop}%
\bibitem [{\citenamefont {Everschor-Sitte}\ \emph {et~al.}(2018)\citenamefont
  {Everschor-Sitte}, \citenamefont {Masell}, \citenamefont {Reeve},\ and\
  \citenamefont {Kl{\"a}ui}}]{skyrmionreview}%
  \BibitemOpen
  \bibfield  {author} {\bibinfo {author} {\bibfnamefont {K.}~\bibnamefont
  {Everschor-Sitte}}, \bibinfo {author} {\bibfnamefont {J.}~\bibnamefont
  {Masell}}, \bibinfo {author} {\bibfnamefont {R.~M.}\ \bibnamefont {Reeve}}, \
  and\ \bibinfo {author} {\bibfnamefont {M.}~\bibnamefont {Kl{\"a}ui}},\
  }\href@noop {} {\bibfield  {journal} {\bibinfo  {journal} {Journal of Applied
  Physics}\ }\textbf {\bibinfo {volume} {124}},\ \bibinfo {pages} {240901}
  (\bibinfo {year} {2018})}\BibitemShut {NoStop}%
\bibitem [{\citenamefont {Jiang}\ \emph {et~al.}(2015)\citenamefont {Jiang},
  \citenamefont {Upadhyaya}, \citenamefont {Zhang}, \citenamefont {Yu},
  \citenamefont {Jungfleisch}, \citenamefont {Fradin}, \citenamefont {Pearson},
  \citenamefont {Tserkovnyak}, \citenamefont {Wang}, \citenamefont {Heinonen},
  \citenamefont {te~Velthuis},\ and\ \citenamefont {Hoffmann}}]{Jiang283}%
  \BibitemOpen
  \bibfield  {author} {\bibinfo {author} {\bibfnamefont {W.}~\bibnamefont
  {Jiang}}, \bibinfo {author} {\bibfnamefont {P.}~\bibnamefont {Upadhyaya}},
  \bibinfo {author} {\bibfnamefont {W.}~\bibnamefont {Zhang}}, \bibinfo
  {author} {\bibfnamefont {G.}~\bibnamefont {Yu}}, \bibinfo {author}
  {\bibfnamefont {M.~B.}\ \bibnamefont {Jungfleisch}}, \bibinfo {author}
  {\bibfnamefont {F.~Y.}\ \bibnamefont {Fradin}}, \bibinfo {author}
  {\bibfnamefont {J.~E.}\ \bibnamefont {Pearson}}, \bibinfo {author}
  {\bibfnamefont {Y.}~\bibnamefont {Tserkovnyak}}, \bibinfo {author}
  {\bibfnamefont {K.~L.}\ \bibnamefont {Wang}}, \bibinfo {author}
  {\bibfnamefont {O.}~\bibnamefont {Heinonen}}, \bibinfo {author}
  {\bibfnamefont {S.~G.~E.}\ \bibnamefont {te~Velthuis}}, \ and\ \bibinfo
  {author} {\bibfnamefont {A.}~\bibnamefont {Hoffmann}},\ }\href@noop {}
  {\bibfield  {journal} {\bibinfo  {journal} {Science}\ }\textbf {\bibinfo
  {volume} {349}},\ \bibinfo {pages} {283} (\bibinfo {year}
  {2015})}\BibitemShut {NoStop}%
\bibitem [{\citenamefont {Ochoa}\ \emph {et~al.}(2016)\citenamefont {Ochoa},
  \citenamefont {Kim},\ and\ \citenamefont {Tserkovnyak}}]{Hectorskyrmion}%
  \BibitemOpen
  \bibfield  {author} {\bibinfo {author} {\bibfnamefont {H.}~\bibnamefont
  {Ochoa}}, \bibinfo {author} {\bibfnamefont {S.~K.}\ \bibnamefont {Kim}}, \
  and\ \bibinfo {author} {\bibfnamefont {Y.}~\bibnamefont {Tserkovnyak}},\
  }\href@noop {} {\bibfield  {journal} {\bibinfo  {journal} {Phys. Rev. B}\
  }\textbf {\bibinfo {volume} {94}},\ \bibinfo {pages} {024431} (\bibinfo
  {year} {2016})}\BibitemShut {NoStop}%
\bibitem [{\citenamefont {Wang}\ \emph {et~al.}(2019)\citenamefont {Wang},
  \citenamefont {Qaiumzadeh},\ and\ \citenamefont
  {Brataas}}]{PhysRevLett.123.147203}%
  \BibitemOpen
  \bibfield  {author} {\bibinfo {author} {\bibfnamefont {X.~S.}\ \bibnamefont
  {Wang}}, \bibinfo {author} {\bibfnamefont {A.}~\bibnamefont {Qaiumzadeh}}, \
  and\ \bibinfo {author} {\bibfnamefont {A.}~\bibnamefont {Brataas}},\
  }\href@noop {} {\bibfield  {journal} {\bibinfo  {journal} {Phys. Rev. Lett.}\
  }\textbf {\bibinfo {volume} {123}},\ \bibinfo {pages} {147203} (\bibinfo
  {year} {2019})}\BibitemShut {NoStop}%
\bibitem [{\citenamefont {Liu}\ \emph {et~al.}(2020)\citenamefont {Liu},
  \citenamefont {Hou}, \citenamefont {Han},\ and\ \citenamefont
  {Zang}}]{Zangprl}%
  \BibitemOpen
  \bibfield  {author} {\bibinfo {author} {\bibfnamefont {Y.}~\bibnamefont
  {Liu}}, \bibinfo {author} {\bibfnamefont {W.}~\bibnamefont {Hou}}, \bibinfo
  {author} {\bibfnamefont {X.}~\bibnamefont {Han}}, \ and\ \bibinfo {author}
  {\bibfnamefont {J.}~\bibnamefont {Zang}},\ }\href@noop {} {\bibfield
  {journal} {\bibinfo  {journal} {Phys. Rev. Lett.}\ }\textbf {\bibinfo
  {volume} {124}},\ \bibinfo {pages} {127204} (\bibinfo {year}
  {2020})}\BibitemShut {NoStop}%
\bibitem [{\citenamefont {Fujishiro}\ \emph {et~al.}(2019)\citenamefont
  {Fujishiro}, \citenamefont {Kanazawa}, \citenamefont {Nakajima},
  \citenamefont {Yu}, \citenamefont {Ohishi}, \citenamefont {Kawamura},
  \citenamefont {Kakurai}, \citenamefont {Arima}, \citenamefont {Mitamura},
  \citenamefont {Miyake}, \citenamefont {Akiba}, \citenamefont {Tokunaga},
  \citenamefont {Matsuo}, \citenamefont {Kindo}, \citenamefont {Koretsune},
  \citenamefont {Arita},\ and\ \citenamefont {Tokura}}]{Fujishiro:2019aa}%
  \BibitemOpen
  \bibfield  {author} {\bibinfo {author} {\bibfnamefont {Y.}~\bibnamefont
  {Fujishiro}}, \bibinfo {author} {\bibfnamefont {N.}~\bibnamefont {Kanazawa}},
  \bibinfo {author} {\bibfnamefont {T.}~\bibnamefont {Nakajima}}, \bibinfo
  {author} {\bibfnamefont {X.~Z.}\ \bibnamefont {Yu}}, \bibinfo {author}
  {\bibfnamefont {K.}~\bibnamefont {Ohishi}}, \bibinfo {author} {\bibfnamefont
  {Y.}~\bibnamefont {Kawamura}}, \bibinfo {author} {\bibfnamefont
  {K.}~\bibnamefont {Kakurai}}, \bibinfo {author} {\bibfnamefont
  {T.}~\bibnamefont {Arima}}, \bibinfo {author} {\bibfnamefont
  {H.}~\bibnamefont {Mitamura}}, \bibinfo {author} {\bibfnamefont
  {A.}~\bibnamefont {Miyake}}, \bibinfo {author} {\bibfnamefont
  {K.}~\bibnamefont {Akiba}}, \bibinfo {author} {\bibfnamefont
  {M.}~\bibnamefont {Tokunaga}}, \bibinfo {author} {\bibfnamefont
  {A.}~\bibnamefont {Matsuo}}, \bibinfo {author} {\bibfnamefont
  {K.}~\bibnamefont {Kindo}}, \bibinfo {author} {\bibfnamefont
  {T.}~\bibnamefont {Koretsune}}, \bibinfo {author} {\bibfnamefont
  {R.}~\bibnamefont {Arita}}, \ and\ \bibinfo {author} {\bibfnamefont
  {Y.}~\bibnamefont {Tokura}},\ }\href@noop {} {\bibfield  {journal} {\bibinfo
  {journal} {Nature Communications}\ }\textbf {\bibinfo {volume} {10}},\
  \bibinfo {pages} {1059} (\bibinfo {year} {2019})}\BibitemShut {NoStop}%
\bibitem [{\citenamefont {Tanigaki}\ \emph {et~al.}(2015)\citenamefont
  {Tanigaki}, \citenamefont {Shibata}, \citenamefont {Kanazawa}, \citenamefont
  {Yu}, \citenamefont {Onose}, \citenamefont {Park}, \citenamefont {Shindo},\
  and\ \citenamefont {Tokura}}]{Tanigaki:2015aa}%
  \BibitemOpen
  \bibfield  {author} {\bibinfo {author} {\bibfnamefont {T.}~\bibnamefont
  {Tanigaki}}, \bibinfo {author} {\bibfnamefont {K.}~\bibnamefont {Shibata}},
  \bibinfo {author} {\bibfnamefont {N.}~\bibnamefont {Kanazawa}}, \bibinfo
  {author} {\bibfnamefont {X.}~\bibnamefont {Yu}}, \bibinfo {author}
  {\bibfnamefont {Y.}~\bibnamefont {Onose}}, \bibinfo {author} {\bibfnamefont
  {H.~S.}\ \bibnamefont {Park}}, \bibinfo {author} {\bibfnamefont
  {D.}~\bibnamefont {Shindo}}, \ and\ \bibinfo {author} {\bibfnamefont
  {Y.}~\bibnamefont {Tokura}},\ }\bibfield  {booktitle} {\emph {\bibinfo
  {booktitle} {Nano Letters}},\ }\href@noop {} {\bibfield  {journal} {\bibinfo
  {journal} {Nano Letters}\ }\textbf {\bibinfo {volume} {15}},\ \bibinfo
  {pages} {5438} (\bibinfo {year} {2015})}\BibitemShut {NoStop}%
\bibitem [{\citenamefont {Kanazawa}\ \emph {et~al.}(2017)\citenamefont
  {Kanazawa}, \citenamefont {Seki},\ and\ \citenamefont
  {Tokura}}]{doi:10.1002/adma.201603227}%
  \BibitemOpen
  \bibfield  {author} {\bibinfo {author} {\bibfnamefont {N.}~\bibnamefont
  {Kanazawa}}, \bibinfo {author} {\bibfnamefont {S.}~\bibnamefont {Seki}}, \
  and\ \bibinfo {author} {\bibfnamefont {Y.}~\bibnamefont {Tokura}},\
  }\href@noop {} {\bibfield  {journal} {\bibinfo  {journal} {Advanced
  Materials}\ }\textbf {\bibinfo {volume} {29}},\ \bibinfo {pages} {1603227}
  (\bibinfo {year} {2017})}\BibitemShut {NoStop}%
\bibitem [{\citenamefont {Nikoli\ifmmode~\acute{c}\else
  \'{c}\fi{}}(2020)}]{PhysRevB.101.115144}%
  \BibitemOpen
  \bibfield  {author} {\bibinfo {author} {\bibfnamefont {P.}~\bibnamefont
  {Nikoli\ifmmode~\acute{c}\else \'{c}\fi{}}},\ }\href@noop {} {\bibfield
  {journal} {\bibinfo  {journal} {Phys. Rev. B}\ }\textbf {\bibinfo {volume}
  {101}},\ \bibinfo {pages} {115144} (\bibinfo {year} {2020})}\BibitemShut
  {NoStop}%
\bibitem [{\citenamefont {Zhang}\ \emph {et~al.}(2016)\citenamefont {Zhang},
  \citenamefont {Mishchenko}, \citenamefont {De~Filippis},\ and\ \citenamefont
  {Nagaosa}}]{XXiao2016prb}%
  \BibitemOpen
  \bibfield  {author} {\bibinfo {author} {\bibfnamefont {X.-X.}\ \bibnamefont
  {Zhang}}, \bibinfo {author} {\bibfnamefont {A.~S.}\ \bibnamefont
  {Mishchenko}}, \bibinfo {author} {\bibfnamefont {G.}~\bibnamefont
  {De~Filippis}}, \ and\ \bibinfo {author} {\bibfnamefont {N.}~\bibnamefont
  {Nagaosa}},\ }\href@noop {} {\bibfield  {journal} {\bibinfo  {journal} {Phys.
  Rev. B}\ }\textbf {\bibinfo {volume} {94}},\ \bibinfo {pages} {174428}
  (\bibinfo {year} {2016})}\BibitemShut {NoStop}%
\bibitem [{\citenamefont {Kanazawa}\ \emph {et~al.}(2016)\citenamefont
  {Kanazawa}, \citenamefont {Nii}, \citenamefont {Zhang}, \citenamefont
  {Mishchenko}, \citenamefont {De~Filippis}, \citenamefont {Kagawa},
  \citenamefont {Iwasa}, \citenamefont {Nagaosa},\ and\ \citenamefont
  {Tokura}}]{Kanazawa:2016ab}%
  \BibitemOpen
  \bibfield  {author} {\bibinfo {author} {\bibfnamefont {N.}~\bibnamefont
  {Kanazawa}}, \bibinfo {author} {\bibfnamefont {Y.}~\bibnamefont {Nii}},
  \bibinfo {author} {\bibfnamefont {X.~X.}\ \bibnamefont {Zhang}}, \bibinfo
  {author} {\bibfnamefont {A.~S.}\ \bibnamefont {Mishchenko}}, \bibinfo
  {author} {\bibfnamefont {G.}~\bibnamefont {De~Filippis}}, \bibinfo {author}
  {\bibfnamefont {F.}~\bibnamefont {Kagawa}}, \bibinfo {author} {\bibfnamefont
  {Y.}~\bibnamefont {Iwasa}}, \bibinfo {author} {\bibfnamefont
  {N.}~\bibnamefont {Nagaosa}}, \ and\ \bibinfo {author} {\bibfnamefont
  {Y.}~\bibnamefont {Tokura}},\ }\href@noop {} {\bibfield  {journal} {\bibinfo
  {journal} {Nature Communications}\ }\textbf {\bibinfo {volume} {7}},\
  \bibinfo {pages} {11622} (\bibinfo {year} {2016})}\BibitemShut {NoStop}%
\bibitem [{\citenamefont {Zarzuela}\ \emph {et~al.}(2019)\citenamefont
  {Zarzuela}, \citenamefont {Ochoa},\ and\ \citenamefont
  {Tserkovnyak}}]{PhysRevB.100.054426}%
  \BibitemOpen
  \bibfield  {author} {\bibinfo {author} {\bibfnamefont {R.}~\bibnamefont
  {Zarzuela}}, \bibinfo {author} {\bibfnamefont {H.}~\bibnamefont {Ochoa}}, \
  and\ \bibinfo {author} {\bibfnamefont {Y.}~\bibnamefont {Tserkovnyak}},\
  }\href@noop {} {\bibfield  {journal} {\bibinfo  {journal} {Phys. Rev. B}\
  }\textbf {\bibinfo {volume} {100}},\ \bibinfo {pages} {054426} (\bibinfo
  {year} {2019})}\BibitemShut {NoStop}%
\bibitem [{\citenamefont {Allwood}\ \emph {et~al.}(2005)\citenamefont
  {Allwood}, \citenamefont {Xiong}, \citenamefont {Faulkner}, \citenamefont
  {Atkinson}, \citenamefont {Petit},\ and\ \citenamefont
  {Cowburn}}]{Allwood1688}%
  \BibitemOpen
  \bibfield  {author} {\bibinfo {author} {\bibfnamefont {D.~A.}\ \bibnamefont
  {Allwood}}, \bibinfo {author} {\bibfnamefont {G.}~\bibnamefont {Xiong}},
  \bibinfo {author} {\bibfnamefont {C.~C.}\ \bibnamefont {Faulkner}}, \bibinfo
  {author} {\bibfnamefont {D.}~\bibnamefont {Atkinson}}, \bibinfo {author}
  {\bibfnamefont {D.}~\bibnamefont {Petit}}, \ and\ \bibinfo {author}
  {\bibfnamefont {R.~P.}\ \bibnamefont {Cowburn}},\ }\href@noop {} {\bibfield
  {journal} {\bibinfo  {journal} {Science}\ }\textbf {\bibinfo {volume}
  {309}},\ \bibinfo {pages} {1688} (\bibinfo {year} {2005})}\BibitemShut
  {NoStop}%
\bibitem [{\citenamefont {Fert}\ \emph {et~al.}(2013)\citenamefont {Fert},
  \citenamefont {Cros},\ and\ \citenamefont {Sampaio}}]{Fert:2013aa}%
  \BibitemOpen
  \bibfield  {author} {\bibinfo {author} {\bibfnamefont {A.}~\bibnamefont
  {Fert}}, \bibinfo {author} {\bibfnamefont {V.}~\bibnamefont {Cros}}, \ and\
  \bibinfo {author} {\bibfnamefont {J.}~\bibnamefont {Sampaio}},\ }\href@noop
  {} {\bibfield  {journal} {\bibinfo  {journal} {Nature Nanotechnology}\
  }\textbf {\bibinfo {volume} {8}},\ \bibinfo {pages} {152} (\bibinfo {year}
  {2013})}\BibitemShut {NoStop}%
\bibitem [{\citenamefont {Parkin}\ \emph {et~al.}(2008)\citenamefont {Parkin},
  \citenamefont {Hayashi},\ and\ \citenamefont {Thomas}}]{Parkin190}%
  \BibitemOpen
  \bibfield  {author} {\bibinfo {author} {\bibfnamefont {S.~S.~P.}\
  \bibnamefont {Parkin}}, \bibinfo {author} {\bibfnamefont {M.}~\bibnamefont
  {Hayashi}}, \ and\ \bibinfo {author} {\bibfnamefont {L.}~\bibnamefont
  {Thomas}},\ }\href@noop {} {\bibfield  {journal} {\bibinfo  {journal}
  {Science}\ }\textbf {\bibinfo {volume} {320}},\ \bibinfo {pages} {190}
  (\bibinfo {year} {2008})}\BibitemShut {NoStop}%
\bibitem [{\citenamefont {Fert}(2008)}]{RevModPhys.80.1517}%
  \BibitemOpen
  \bibfield  {author} {\bibinfo {author} {\bibfnamefont {A.}~\bibnamefont
  {Fert}},\ }\href@noop {} {\bibfield  {journal} {\bibinfo  {journal} {Rev.
  Mod. Phys.}\ }\textbf {\bibinfo {volume} {80}},\ \bibinfo {pages} {1517}
  (\bibinfo {year} {2008})}\BibitemShut {NoStop}%
\bibitem [{\citenamefont {Tserkovnyak}\ and\ \citenamefont
  {Xiao}(2018)}]{PhysRevLett.121.127701}%
  \BibitemOpen
  \bibfield  {author} {\bibinfo {author} {\bibfnamefont {Y.}~\bibnamefont
  {Tserkovnyak}}\ and\ \bibinfo {author} {\bibfnamefont {J.}~\bibnamefont
  {Xiao}},\ }\href@noop {} {\bibfield  {journal} {\bibinfo  {journal} {Phys.
  Rev. Lett.}\ }\textbf {\bibinfo {volume} {121}},\ \bibinfo {pages} {127701}
  (\bibinfo {year} {2018})}\BibitemShut {NoStop}%
\bibitem [{\citenamefont {{Jones}}\ \emph {et~al.}(2020)\citenamefont
  {{Jones}}, \citenamefont {{Zou}}, \citenamefont {{Zhang}},\ and\
  \citenamefont {{Tserkovnyak}}}]{energystorage}%
  \BibitemOpen
  \bibfield  {author} {\bibinfo {author} {\bibfnamefont {D.}~\bibnamefont
  {{Jones}}}, \bibinfo {author} {\bibfnamefont {J.}~\bibnamefont {{Zou}}},
  \bibinfo {author} {\bibfnamefont {S.}~\bibnamefont {{Zhang}}}, \ and\
  \bibinfo {author} {\bibfnamefont {Y.}~\bibnamefont {{Tserkovnyak}}},\
  }\href@noop {} {\bibfield  {journal} {\bibinfo  {journal} {arXiv e-prints}\
  ,\ \bibinfo {eid} {arXiv:2003.12121}} (\bibinfo {year} {2020})}\BibitemShut
  {NoStop}%
\bibitem [{\citenamefont {Nakahara}(2003)}]{nkhr}%
  \BibitemOpen
  \bibfield  {author} {\bibinfo {author} {\bibfnamefont {M.}~\bibnamefont
  {Nakahara}},\ }\href@noop {} {\emph {\bibinfo {title} {Geometry, Topology and
  Physics}}},\ \bibinfo {edition} {2nd}\ ed.\ (\bibinfo  {publisher} {CRC
  Press},\ \bibinfo {year} {2003})\BibitemShut {NoStop}%
\bibitem [{sm()}]{sm}%
  \BibitemOpen
  \href@noop {} {}\bibinfo {note} {See Supplemental Material for
  (i) proof of the topological conservation law, (ii) topological conservation
  law of hedgehogs on a lattice, (iii) electric charge distribution for global
  dynamics, and (iv) hedgehog conductivity from collective variable
  approach, which includes Ref. \cite{quantumvortex, Hectorskyrmion, Stoneskyrmion, 1905AnP...322..549E, xiaogangwenqft}.}\BibitemShut {Stop}%
  \bibitem [{\citenamefont {Stone}(1996)}]{Stoneskyrmion}%
  \BibitemOpen
  \bibfield  {author} {\bibinfo {author} {\bibfnamefont {M.}~\bibnamefont
  {Stone}},\ }\href@noop {} {\bibfield  {journal} {\bibinfo  {journal} {Phys.
  Rev. B}\ }\textbf {\bibinfo {volume} {53}},\ \bibinfo {pages} {16573}
  (\bibinfo {year} {1996})}\BibitemShut {NoStop}%
\bibitem [{\citenamefont {{Einstein}}(1905)}]{1905AnP...322..549E}%
  \BibitemOpen
  \bibfield  {author} {\bibinfo {author} {\bibfnamefont {A.}~\bibnamefont
  {{Einstein}}},\ }\href@noop {} {\bibfield  {journal} {\bibinfo  {journal}
  {Annalen der Physik}\ }\textbf {\bibinfo {volume} {322}},\ \bibinfo {pages}
  {549} (\bibinfo {year} {1905})}\BibitemShut {NoStop}%
\bibitem [{\citenamefont {Wen}(2007)}]{xiaogangwenqft}%
  \BibitemOpen
  \bibfield  {author} {\bibinfo {author} {\bibfnamefont {X.-G.}\ \bibnamefont
  {Wen}},\ }\href@noop {} {\emph {\bibinfo {title} {Quantum Field Theory of
  Many-body Systems: From the Origin of Sound to an Origin of Light and
  Electrons}}}\ (\bibinfo  {publisher} {Oxford University Press},\ \bibinfo
  {year} {2007})\BibitemShut {NoStop}%
\bibitem [{\citenamefont {Milde}\ \emph {et~al.}(2013)\citenamefont {Milde},
  \citenamefont {K{\"o}hler}, \citenamefont {Seidel}, \citenamefont {Eng},
  \citenamefont {Bauer}, \citenamefont {Chacon}, \citenamefont {Kindervater},
  \citenamefont {M{\"u}hlbauer}, \citenamefont {Pfleiderer}, \citenamefont
  {Buhrandt}, \citenamefont {Sch{\"u}tte},\ and\ \citenamefont
  {Rosch}}]{Milde1076}%
  \BibitemOpen
  \bibfield  {author} {\bibinfo {author} {\bibfnamefont {P.}~\bibnamefont
  {Milde}}, \bibinfo {author} {\bibfnamefont {D.}~\bibnamefont {K{\"o}hler}},
  \bibinfo {author} {\bibfnamefont {J.}~\bibnamefont {Seidel}}, \bibinfo
  {author} {\bibfnamefont {L.~M.}\ \bibnamefont {Eng}}, \bibinfo {author}
  {\bibfnamefont {A.}~\bibnamefont {Bauer}}, \bibinfo {author} {\bibfnamefont
  {A.}~\bibnamefont {Chacon}}, \bibinfo {author} {\bibfnamefont
  {J.}~\bibnamefont {Kindervater}}, \bibinfo {author} {\bibfnamefont
  {S.}~\bibnamefont {M{\"u}hlbauer}}, \bibinfo {author} {\bibfnamefont
  {C.}~\bibnamefont {Pfleiderer}}, \bibinfo {author} {\bibfnamefont
  {S.}~\bibnamefont {Buhrandt}}, \bibinfo {author} {\bibfnamefont
  {C.}~\bibnamefont {Sch{\"u}tte}}, \ and\ \bibinfo {author} {\bibfnamefont
  {A.}~\bibnamefont {Rosch}},\ }\href@noop {} {\bibfield  {journal} {\bibinfo
  {journal} {Science}\ }\textbf {\bibinfo {volume} {340}},\ \bibinfo {pages}
  {1076} (\bibinfo {year} {2013})}\BibitemShut {NoStop}%
\bibitem [{\citenamefont {Volovik}(1987)}]{linearmomentum}%
  \BibitemOpen
  \bibfield  {author} {\bibinfo {author} {\bibfnamefont {G.~E.}\ \bibnamefont
  {Volovik}},\ }\href@noop {} {\bibfield  {journal} {\bibinfo  {journal} {J.
  Phys. C: Solid State Phys.}\ }\textbf {\bibinfo {volume} {20}} (\bibinfo
  {year} {1987})}\BibitemShut {NoStop}%
\bibitem [{\citenamefont {Nagaosa}\ and\ \citenamefont
  {Tokura}(2012)}]{Nagaosa_2012}%
  \BibitemOpen
  \bibfield  {author} {\bibinfo {author} {\bibfnamefont {N.}~\bibnamefont
  {Nagaosa}}\ and\ \bibinfo {author} {\bibfnamefont {Y.}~\bibnamefont
  {Tokura}},\ }\href@noop {} {\bibfield  {journal} {\bibinfo  {journal}
  {Physica Scripta}\ }\textbf {\bibinfo {volume} {T146}},\ \bibinfo {pages}
  {014020} (\bibinfo {year} {2012})}\BibitemShut {NoStop}%
\bibitem [{\citenamefont {Wong}\ and\ \citenamefont
  {Tserkovnyak}(2009)}]{PhysRevB.80.184411}%
  \BibitemOpen
  \bibfield  {author} {\bibinfo {author} {\bibfnamefont {C.~H.}\ \bibnamefont
  {Wong}}\ and\ \bibinfo {author} {\bibfnamefont {Y.}~\bibnamefont
  {Tserkovnyak}},\ }\href@noop {} {\bibfield  {journal} {\bibinfo  {journal}
  {Phys. Rev. B}\ }\textbf {\bibinfo {volume} {80}},\ \bibinfo {pages} {184411}
  (\bibinfo {year} {2009})}\BibitemShut {NoStop}%
\bibitem [{\citenamefont {Tatara}(2019)}]{TATARA2019208}%
  \BibitemOpen
  \bibfield  {author} {\bibinfo {author} {\bibfnamefont {G.}~\bibnamefont
  {Tatara}},\ }\href@noop {} {\bibfield  {journal} {\bibinfo  {journal}
  {Physica E: Low-dimensional Systems and Nanostructures}\ }\textbf {\bibinfo
  {volume} {106}},\ \bibinfo {pages} {208 } (\bibinfo {year}
  {2019})}\BibitemShut {NoStop}%
\bibitem [{\citenamefont {Schulz}\ \emph {et~al.}(2012)\citenamefont {Schulz},
  \citenamefont {Ritz}, \citenamefont {Bauer}, \citenamefont {Halder},
  \citenamefont {Wagner}, \citenamefont {Franz}, \citenamefont {Pfleiderer},
  \citenamefont {Everschor}, \citenamefont {Garst},\ and\ \citenamefont
  {Rosch}}]{Schulz:2012ab}%
  \BibitemOpen
  \bibfield  {author} {\bibinfo {author} {\bibfnamefont {T.}~\bibnamefont
  {Schulz}}, \bibinfo {author} {\bibfnamefont {R.}~\bibnamefont {Ritz}},
  \bibinfo {author} {\bibfnamefont {A.}~\bibnamefont {Bauer}}, \bibinfo
  {author} {\bibfnamefont {M.}~\bibnamefont {Halder}}, \bibinfo {author}
  {\bibfnamefont {M.}~\bibnamefont {Wagner}}, \bibinfo {author} {\bibfnamefont
  {C.}~\bibnamefont {Franz}}, \bibinfo {author} {\bibfnamefont
  {C.}~\bibnamefont {Pfleiderer}}, \bibinfo {author} {\bibfnamefont
  {K.}~\bibnamefont {Everschor}}, \bibinfo {author} {\bibfnamefont
  {M.}~\bibnamefont {Garst}}, \ and\ \bibinfo {author} {\bibfnamefont
  {A.}~\bibnamefont {Rosch}},\ }\href@noop {} {\bibfield  {journal} {\bibinfo
  {journal} {Nature Physics}\ }\textbf {\bibinfo {volume} {8}},\ \bibinfo
  {pages} {301} (\bibinfo {year} {2012})}\BibitemShut {NoStop}%
\bibitem [{\citenamefont {D\'{\i}az}\ \emph {et~al.}(2019)\citenamefont
  {D\'{\i}az}, \citenamefont {Klinovaja},\ and\ \citenamefont
  {Loss}}]{PhysRevLett.122.187203}%
  \BibitemOpen
  \bibfield  {author} {\bibinfo {author} {\bibfnamefont {S.~A.}\ \bibnamefont
  {D\'{\i}az}}, \bibinfo {author} {\bibfnamefont {J.}~\bibnamefont
  {Klinovaja}}, \ and\ \bibinfo {author} {\bibfnamefont {D.}~\bibnamefont
  {Loss}},\ }\href@noop {} {\bibfield  {journal} {\bibinfo  {journal} {Phys.
  Rev. Lett.}\ }\textbf {\bibinfo {volume} {122}},\ \bibinfo {pages} {187203}
  (\bibinfo {year} {2019})}\BibitemShut {NoStop}%
\bibitem [{\citenamefont {van Hoogdalem}\ \emph {et~al.}(2013)\citenamefont
  {van Hoogdalem}, \citenamefont {Tserkovnyak},\ and\ \citenamefont
  {Loss}}]{PhysRevB.87.024402}%
  \BibitemOpen
  \bibfield  {author} {\bibinfo {author} {\bibfnamefont {K.~A.}\ \bibnamefont
  {van Hoogdalem}}, \bibinfo {author} {\bibfnamefont {Y.}~\bibnamefont
  {Tserkovnyak}}, \ and\ \bibinfo {author} {\bibfnamefont {D.}~\bibnamefont
  {Loss}},\ }\href@noop {} {\bibfield  {journal} {\bibinfo  {journal} {Phys.
  Rev. B}\ }\textbf {\bibinfo {volume} {87}},\ \bibinfo {pages} {024402}
  (\bibinfo {year} {2013})}\BibitemShut {NoStop}%
\bibitem [{\citenamefont {Coleman}(1988)}]{aspectofsymmetry}%
  \BibitemOpen
  \bibfield  {author} {\bibinfo {author} {\bibfnamefont {S.}~\bibnamefont
  {Coleman}},\ }\href@noop {} {\emph {\bibinfo {title} {Aspects of Symmetry:
  Selected Erice Lectures}}}\ (\bibinfo  {publisher} {Cambridge University
  Press},\ \bibinfo {year} {1988})\BibitemShut {NoStop}%
\bibitem [{\citenamefont {Polyakov}(1987)}]{gaugefieldstring}%
  \BibitemOpen
  \bibfield  {author} {\bibinfo {author} {\bibfnamefont {A.~M.}\ \bibnamefont
  {Polyakov}},\ }\href@noop {} {\emph {\bibinfo {title} {Gauge fields and
  strings}}}\ (\bibinfo  {publisher} {CRC Press},\ \bibinfo {year} {September
  14, 1987})\BibitemShut {NoStop}%
\bibitem [{\citenamefont {Brezis}\ \emph {et~al.}(1986)\citenamefont {Brezis},
  \citenamefont {Coron},\ and\ \citenamefont {Lieb}}]{brezis1986}%
  \BibitemOpen
  \bibfield  {author} {\bibinfo {author} {\bibfnamefont {H.}~\bibnamefont
  {Brezis}}, \bibinfo {author} {\bibfnamefont {J.-M.}\ \bibnamefont {Coron}}, \
  and\ \bibinfo {author} {\bibfnamefont {E.~H.}\ \bibnamefont {Lieb}},\
  }\href@noop {} {\bibfield  {journal} {\bibinfo  {journal} {Comm. Math.
  Phys.}\ }\textbf {\bibinfo {volume} {107}},\ \bibinfo {pages} {649} (\bibinfo
  {year} {1986})}\BibitemShut {NoStop}%
\bibitem [{\citenamefont {Kogut}(1983)}]{RevModPhys.55.775}%
  \BibitemOpen
  \bibfield  {author} {\bibinfo {author} {\bibfnamefont {J.~B.}\ \bibnamefont
  {Kogut}},\ }\href@noop {} {\bibfield  {journal} {\bibinfo  {journal} {Rev.
  Mod. Phys.}\ }\textbf {\bibinfo {volume} {55}},\ \bibinfo {pages} {775}
  (\bibinfo {year} {1983})}\BibitemShut {NoStop}%
\bibitem [{\citenamefont {Kadanoff}(1977)}]{RevModPhys.49.267}%
  \BibitemOpen
  \bibfield  {author} {\bibinfo {author} {\bibfnamefont {L.~P.}\ \bibnamefont
  {Kadanoff}},\ }\href@noop {} {\bibfield  {journal} {\bibinfo  {journal} {Rev.
  Mod. Phys.}\ }\textbf {\bibinfo {volume} {49}},\ \bibinfo {pages} {267}
  (\bibinfo {year} {1977})}\BibitemShut {NoStop}%
\bibitem [{\citenamefont {McLerran}(1986)}]{RevModPhys.58.1021}%
  \BibitemOpen
  \bibfield  {author} {\bibinfo {author} {\bibfnamefont {L.}~\bibnamefont
  {McLerran}},\ }\href@noop {} {\bibfield  {journal} {\bibinfo  {journal} {Rev.
  Mod. Phys.}\ }\textbf {\bibinfo {volume} {58}},\ \bibinfo {pages} {1021}
  (\bibinfo {year} {1986})}\BibitemShut {NoStop}%
\bibitem [{\citenamefont {Bohr}\ and\ \citenamefont
  {Nielsen}(1977)}]{quarksoup}%
  \BibitemOpen
  \bibfield  {author} {\bibinfo {author} {\bibfnamefont {H.}~\bibnamefont
  {Bohr}}\ and\ \bibinfo {author} {\bibfnamefont {H.}~\bibnamefont {Nielsen}},\
  }\href@noop {} {\bibfield  {journal} {\bibinfo  {journal} {Nuclear Physics
  B}\ }\textbf {\bibinfo {volume} {128}},\ \bibinfo {pages} {275 } (\bibinfo
  {year} {1977})}\BibitemShut {NoStop}%
\bibitem [{\citenamefont {Cassing}\ and\ \citenamefont
  {Bratkovskaya}(2008)}]{PhysRevC.78.034919}%
  \BibitemOpen
  \bibfield  {author} {\bibinfo {author} {\bibfnamefont {W.}~\bibnamefont
  {Cassing}}\ and\ \bibinfo {author} {\bibfnamefont {E.~L.}\ \bibnamefont
  {Bratkovskaya}},\ }\href@noop {} {\bibfield  {journal} {\bibinfo  {journal}
  {Phys. Rev. C}\ }\textbf {\bibinfo {volume} {78}},\ \bibinfo {pages} {034919}
  (\bibinfo {year} {2008})}\BibitemShut {NoStop}%
\bibitem [{\citenamefont {Peshier}\ and\ \citenamefont
  {Cassing}(2005)}]{PhysRevLett.94.172301}%
  \BibitemOpen
  \bibfield  {author} {\bibinfo {author} {\bibfnamefont {A.}~\bibnamefont
  {Peshier}}\ and\ \bibinfo {author} {\bibfnamefont {W.}~\bibnamefont
  {Cassing}},\ }\href@noop {} {\bibfield  {journal} {\bibinfo  {journal} {Phys.
  Rev. Lett.}\ }\textbf {\bibinfo {volume} {94}},\ \bibinfo {pages} {172301}
  (\bibinfo {year} {2005})}\BibitemShut {NoStop}%
\bibitem [{\citenamefont {Nielsen}\ and\ \citenamefont
  {Olesen}(1973)}]{Nielsen:1973cs}%
  \BibitemOpen
  \bibfield  {author} {\bibinfo {author} {\bibfnamefont {H.~B.}\ \bibnamefont
  {Nielsen}}\ and\ \bibinfo {author} {\bibfnamefont {P.}~\bibnamefont
  {Olesen}},\ }\href@noop {} {\bibfield  {journal} {\bibinfo  {journal} {Nucl.
  Phys.}\ }\textbf {\bibinfo {volume} {B61}},\ \bibinfo {pages} {45} (\bibinfo
  {year} {1973})},\ \bibinfo {note} {[,302(1973)]}\BibitemShut {NoStop}%
\bibitem [{\citenamefont {Nambu}(1974)}]{Nambuconfine}%
  \BibitemOpen
  \bibfield  {author} {\bibinfo {author} {\bibfnamefont {Y.}~\bibnamefont
  {Nambu}},\ }\href@noop {} {\bibfield  {journal} {\bibinfo  {journal} {Phys.
  Rev. D}\ }\textbf {\bibinfo {volume} {10}},\ \bibinfo {pages} {4262}
  (\bibinfo {year} {1974})}\BibitemShut {NoStop}%
\bibitem [{\citenamefont {Hooft}(1974)}]{HOOFT1974276}%
  \BibitemOpen
  \bibfield  {author} {\bibinfo {author} {\bibfnamefont {G.}~\bibnamefont
  {Hooft}},\ }\href@noop {} {\bibfield  {journal} {\bibinfo  {journal} {Nuclear
  Physics B}\ }\textbf {\bibinfo {volume} {79}},\ \bibinfo {pages} {276 }
  (\bibinfo {year} {1974})}\BibitemShut {NoStop}%
\bibitem [{\citenamefont {Ball}\ and\ \citenamefont
  {Caticha}(1988)}]{PhysRevD.37.524}%
  \BibitemOpen
  \bibfield  {author} {\bibinfo {author} {\bibfnamefont {J.~S.}\ \bibnamefont
  {Ball}}\ and\ \bibinfo {author} {\bibfnamefont {A.}~\bibnamefont {Caticha}},\
  }\href@noop {} {\bibfield  {journal} {\bibinfo  {journal} {Phys. Rev. D}\
  }\textbf {\bibinfo {volume} {37}},\ \bibinfo {pages} {524} (\bibinfo {year}
  {1988})}\BibitemShut {NoStop}%
\bibitem [{\citenamefont {Cardoso}\ \emph {et~al.}(2008)\citenamefont
  {Cardoso}, \citenamefont {Bicudo},\ and\ \citenamefont
  {Sacramento}}]{CARDOSO2008337}%
  \BibitemOpen
  \bibfield  {author} {\bibinfo {author} {\bibfnamefont {M.}~\bibnamefont
  {Cardoso}}, \bibinfo {author} {\bibfnamefont {P.}~\bibnamefont {Bicudo}}, \
  and\ \bibinfo {author} {\bibfnamefont {P.~D.}\ \bibnamefont {Sacramento}},\
  }\href@noop {} {\bibfield  {journal} {\bibinfo  {journal} {Annals of
  Physics}\ }\textbf {\bibinfo {volume} {323}},\ \bibinfo {pages} {337 }
  (\bibinfo {year} {2008})}\BibitemShut {NoStop}%
\bibitem [{\citenamefont {Akagi}\ \emph {et~al.}(2012)\citenamefont {Akagi},
  \citenamefont {Udagawa},\ and\ \citenamefont {Motome}}]{higherorder}%
  \BibitemOpen
  \bibfield  {author} {\bibinfo {author} {\bibfnamefont {Y.}~\bibnamefont
  {Akagi}}, \bibinfo {author} {\bibfnamefont {M.}~\bibnamefont {Udagawa}}, \
  and\ \bibinfo {author} {\bibfnamefont {Y.}~\bibnamefont {Motome}},\
  }\href@noop {} {\bibfield  {journal} {\bibinfo  {journal} {Phys. Rev. Lett.}\
  }\textbf {\bibinfo {volume} {108}},\ \bibinfo {pages} {096401} (\bibinfo
  {year} {2012})}\BibitemShut {NoStop}%
\bibitem [{\citenamefont {Okumura}\ \emph {et~al.}(2020)\citenamefont
  {Okumura}, \citenamefont {Hayami}, \citenamefont {Kato},\ and\ \citenamefont
  {Motome}}]{PhysRevB.101.144416}%
  \BibitemOpen
  \bibfield  {author} {\bibinfo {author} {\bibfnamefont {S.}~\bibnamefont
  {Okumura}}, \bibinfo {author} {\bibfnamefont {S.}~\bibnamefont {Hayami}},
  \bibinfo {author} {\bibfnamefont {Y.}~\bibnamefont {Kato}}, \ and\ \bibinfo
  {author} {\bibfnamefont {Y.}~\bibnamefont {Motome}},\ }\href@noop {}
  {\bibfield  {journal} {\bibinfo  {journal} {Phys. Rev. B}\ }\textbf {\bibinfo
  {volume} {101}},\ \bibinfo {pages} {144416} (\bibinfo {year}
  {2020})}\BibitemShut {NoStop}%
\bibitem [{\citenamefont {Sch\"utte}\ and\ \citenamefont
  {Rosch}(2014)}]{PhysRevB.90.174432}%
  \BibitemOpen
  \bibfield  {author} {\bibinfo {author} {\bibfnamefont {C.}~\bibnamefont
  {Sch\"utte}}\ and\ \bibinfo {author} {\bibfnamefont {A.}~\bibnamefont
  {Rosch}},\ }\href@noop {} {\bibfield  {journal} {\bibinfo  {journal} {Phys.
  Rev. B}\ }\textbf {\bibinfo {volume} {90}},\ \bibinfo {pages} {174432}
  (\bibinfo {year} {2014})}\BibitemShut {NoStop}%
\bibitem [{\citenamefont {Takei}\ and\ \citenamefont
  {Tserkovnyak}(2014)}]{YSST}%
  \BibitemOpen
  \bibfield  {author} {\bibinfo {author} {\bibfnamefont {S.}~\bibnamefont
  {Takei}}\ and\ \bibinfo {author} {\bibfnamefont {Y.}~\bibnamefont
  {Tserkovnyak}},\ }\href@noop {} {\bibfield  {journal} {\bibinfo  {journal}
  {Phys. Rev. Lett.}\ }\textbf {\bibinfo {volume} {112}},\ \bibinfo {pages}
  {227201} (\bibinfo {year} {2014})}\BibitemShut {NoStop}%
\bibitem [{\citenamefont {Kim}\ \emph {et~al.}(2018)\citenamefont {Kim},
  \citenamefont {Myers},\ and\ \citenamefont {Tserkovnyak}}]{Sekwonvor}%
  \BibitemOpen
  \bibfield  {author} {\bibinfo {author} {\bibfnamefont {S.~K.}\ \bibnamefont
  {Kim}}, \bibinfo {author} {\bibfnamefont {R.}~\bibnamefont {Myers}}, \ and\
  \bibinfo {author} {\bibfnamefont {Y.}~\bibnamefont {Tserkovnyak}},\
  }\href@noop {} {\bibfield  {journal} {\bibinfo  {journal} {Phys. Rev. Lett.}\
  }\textbf {\bibinfo {volume} {121}},\ \bibinfo {pages} {187203} (\bibinfo
  {year} {2018})}\BibitemShut {NoStop}%
\bibitem [{\citenamefont {H\"anggi}\ \emph {et~al.}(1990)\citenamefont
  {H\"anggi}, \citenamefont {Talkner},\ and\ \citenamefont
  {Borkovec}}]{reactionrate}%
  \BibitemOpen
  \bibfield  {author} {\bibinfo {author} {\bibfnamefont {P.}~\bibnamefont
  {H\"anggi}}, \bibinfo {author} {\bibfnamefont {P.}~\bibnamefont {Talkner}}, \
  and\ \bibinfo {author} {\bibfnamefont {M.}~\bibnamefont {Borkovec}},\
  }\href@noop {} {\bibfield  {journal} {\bibinfo  {journal} {Rev. Mod. Phys.}\
  }\textbf {\bibinfo {volume} {62}},\ \bibinfo {pages} {251} (\bibinfo {year}
  {1990})}\BibitemShut {NoStop}%
\bibitem [{\citenamefont {Tserkovnyak}\ and\ \citenamefont
  {Mecklenburg}(2008)}]{PhysRevB.77.134407}%
  \BibitemOpen
  \bibfield  {author} {\bibinfo {author} {\bibfnamefont {Y.}~\bibnamefont
  {Tserkovnyak}}\ and\ \bibinfo {author} {\bibfnamefont {M.}~\bibnamefont
  {Mecklenburg}},\ }\href@noop {} {\bibfield  {journal} {\bibinfo  {journal}
  {Phys. Rev. B}\ }\textbf {\bibinfo {volume} {77}},\ \bibinfo {pages} {134407}
  (\bibinfo {year} {2008})}\BibitemShut {NoStop}%
\bibitem [{\citenamefont {Onsager}(1931)}]{PhysRev.37.405}%
  \BibitemOpen
  \bibfield  {author} {\bibinfo {author} {\bibfnamefont {L.}~\bibnamefont
  {Onsager}},\ }\href@noop {} {\bibfield  {journal} {\bibinfo  {journal} {Phys.
  Rev.}\ }\textbf {\bibinfo {volume} {37}},\ \bibinfo {pages} {405} (\bibinfo
  {year} {1931})}\BibitemShut {NoStop}%
\bibitem [{\citenamefont {Donnelly}\ \emph {et~al.}(2017)\citenamefont
  {Donnelly}, \citenamefont {Guizar-Sicairos}, \citenamefont {Scagnoli},
  \citenamefont {Gliga}, \citenamefont {Holler}, \citenamefont {Raabe},\ and\
  \citenamefont {Heyderman}}]{Donnelly:2017aa}%
  \BibitemOpen
  \bibfield  {author} {\bibinfo {author} {\bibfnamefont {C.}~\bibnamefont
  {Donnelly}}, \bibinfo {author} {\bibfnamefont {M.}~\bibnamefont
  {Guizar-Sicairos}}, \bibinfo {author} {\bibfnamefont {V.}~\bibnamefont
  {Scagnoli}}, \bibinfo {author} {\bibfnamefont {S.}~\bibnamefont {Gliga}},
  \bibinfo {author} {\bibfnamefont {M.}~\bibnamefont {Holler}}, \bibinfo
  {author} {\bibfnamefont {J.}~\bibnamefont {Raabe}}, \ and\ \bibinfo {author}
  {\bibfnamefont {L.~J.}\ \bibnamefont {Heyderman}},\ }\href@noop {} {\bibfield
   {journal} {\bibinfo  {journal} {Nature}\ }\textbf {\bibinfo {volume}
  {547}},\ \bibinfo {pages} {328} (\bibinfo {year} {2017})}\BibitemShut
  {NoStop}%
\bibitem [{\citenamefont {Donnelly}\ \emph {et~al.}(2020)\citenamefont
  {Donnelly}, \citenamefont {Finizio}, \citenamefont {Gliga}, \citenamefont
  {Holler}, \citenamefont {Hrabec}, \citenamefont {Odstr{\v c}il},
  \citenamefont {Mayr}, \citenamefont {Scagnoli}, \citenamefont {Heyderman},
  \citenamefont {Guizar-Sicairos},\ and\ \citenamefont
  {Raabe}}]{Donnelly:2020aa}%
  \BibitemOpen
  \bibfield  {author} {\bibinfo {author} {\bibfnamefont {C.}~\bibnamefont
  {Donnelly}}, \bibinfo {author} {\bibfnamefont {S.}~\bibnamefont {Finizio}},
  \bibinfo {author} {\bibfnamefont {S.}~\bibnamefont {Gliga}}, \bibinfo
  {author} {\bibfnamefont {M.}~\bibnamefont {Holler}}, \bibinfo {author}
  {\bibfnamefont {A.}~\bibnamefont {Hrabec}}, \bibinfo {author} {\bibfnamefont
  {M.}~\bibnamefont {Odstr{\v c}il}}, \bibinfo {author} {\bibfnamefont
  {S.}~\bibnamefont {Mayr}}, \bibinfo {author} {\bibfnamefont {V.}~\bibnamefont
  {Scagnoli}}, \bibinfo {author} {\bibfnamefont {L.~J.}\ \bibnamefont
  {Heyderman}}, \bibinfo {author} {\bibfnamefont {M.}~\bibnamefont
  {Guizar-Sicairos}}, \ and\ \bibinfo {author} {\bibfnamefont {J.}~\bibnamefont
  {Raabe}},\ }\href@noop {} {\bibfield  {journal} {\bibinfo  {journal} {Nature
  Nanotechnology}\ }\textbf {\bibinfo {volume} {15}},\ \bibinfo {pages} {356}
  (\bibinfo {year} {2020})}\BibitemShut {NoStop}%
\bibitem [{\citenamefont {Romming}\ \emph {et~al.}(2013)\citenamefont
  {Romming}, \citenamefont {Hanneken}, \citenamefont {Menzel}, \citenamefont
  {Bickel}, \citenamefont {Wolter}, \citenamefont {von Bergmann}, \citenamefont
  {Kubetzka},\ and\ \citenamefont {Wiesendanger}}]{Romming636}%
  \BibitemOpen
  \bibfield  {author} {\bibinfo {author} {\bibfnamefont {N.}~\bibnamefont
  {Romming}}, \bibinfo {author} {\bibfnamefont {C.}~\bibnamefont {Hanneken}},
  \bibinfo {author} {\bibfnamefont {M.}~\bibnamefont {Menzel}}, \bibinfo
  {author} {\bibfnamefont {J.~E.}\ \bibnamefont {Bickel}}, \bibinfo {author}
  {\bibfnamefont {B.}~\bibnamefont {Wolter}}, \bibinfo {author} {\bibfnamefont
  {K.}~\bibnamefont {von Bergmann}}, \bibinfo {author} {\bibfnamefont
  {A.}~\bibnamefont {Kubetzka}}, \ and\ \bibinfo {author} {\bibfnamefont
  {R.}~\bibnamefont {Wiesendanger}},\ }\href@noop {} {\bibfield  {journal}
  {\bibinfo  {journal} {Science}\ }\textbf {\bibinfo {volume} {341}},\ \bibinfo
  {pages} {636} (\bibinfo {year} {2013})}\BibitemShut {NoStop}%
\bibitem [{\citenamefont {Chen}\ and\ \citenamefont
  {Schmid}(2015)}]{doi:10.1002/adma.201500160}%
  \BibitemOpen
  \bibfield  {author} {\bibinfo {author} {\bibfnamefont {G.}~\bibnamefont
  {Chen}}\ and\ \bibinfo {author} {\bibfnamefont {A.~K.}\ \bibnamefont
  {Schmid}},\ }\href@noop {} {\bibfield  {journal} {\bibinfo  {journal}
  {Advanced Materials}\ }\textbf {\bibinfo {volume} {27}},\ \bibinfo {pages}
  {5738} (\bibinfo {year} {2015})}\BibitemShut {NoStop}%
\bibitem [{\citenamefont {Rougemaille}\ and\ \citenamefont
  {Schmid}(2010)}]{rougemaille2010magnetic}%
  \BibitemOpen
  \bibfield  {author} {\bibinfo {author} {\bibfnamefont {N.}~\bibnamefont
  {Rougemaille}}\ and\ \bibinfo {author} {\bibfnamefont {A.}~\bibnamefont
  {Schmid}},\ }\href@noop {} {\bibfield  {journal} {\bibinfo  {journal} {The
  European Physical Journal-Applied Physics}\ }\textbf {\bibinfo {volume} {50}}
  (\bibinfo {year} {2010})}\BibitemShut {NoStop}%
\bibitem [{\citenamefont {Ochoa}\ \emph {et~al.}(2017)\citenamefont {Ochoa},
  \citenamefont {Kim}, \citenamefont {Tchernyshyov},\ and\ \citenamefont
  {Tserkovnyak}}]{PhysRevB.96.020410}%
  \BibitemOpen
  \bibfield  {author} {\bibinfo {author} {\bibfnamefont {H.}~\bibnamefont
  {Ochoa}}, \bibinfo {author} {\bibfnamefont {S.~K.}\ \bibnamefont {Kim}},
  \bibinfo {author} {\bibfnamefont {O.}~\bibnamefont {Tchernyshyov}}, \ and\
  \bibinfo {author} {\bibfnamefont {Y.}~\bibnamefont {Tserkovnyak}},\
  }\href@noop {} {\bibfield  {journal} {\bibinfo  {journal} {Phys. Rev. B}\
  }\textbf {\bibinfo {volume} {96}},\ \bibinfo {pages} {020410} (\bibinfo
  {year} {2017})}\BibitemShut {NoStop}%
\bibitem [{\citenamefont {Stier}\ \emph {et~al.}(2017)\citenamefont {Stier},
  \citenamefont {H\"ausler}, \citenamefont {Posske}, \citenamefont {Gurski},\
  and\ \citenamefont {Thorwart}}]{StierPRL2017}%
  \BibitemOpen
  \bibfield  {author} {\bibinfo {author} {\bibfnamefont {M.}~\bibnamefont
  {Stier}}, \bibinfo {author} {\bibfnamefont {W.}~\bibnamefont {H\"ausler}},
  \bibinfo {author} {\bibfnamefont {T.}~\bibnamefont {Posske}}, \bibinfo
  {author} {\bibfnamefont {G.}~\bibnamefont {Gurski}}, \ and\ \bibinfo {author}
  {\bibfnamefont {M.}~\bibnamefont {Thorwart}},\ }\href@noop {} {\bibfield
  {journal} {\bibinfo  {journal} {Phys. Rev. Lett.}\ }\textbf {\bibinfo
  {volume} {118}},\ \bibinfo {pages} {267203} (\bibinfo {year}
  {2017})}\BibitemShut {NoStop}%
\bibitem [{\citenamefont {Everschor-Sitte}\ \emph {et~al.}(2017)\citenamefont
  {Everschor-Sitte}, \citenamefont {Sitte}, \citenamefont {Valet},
  \citenamefont {Abanov},\ and\ \citenamefont {Sinova}}]{Everschor_Sitte2017}%
  \BibitemOpen
  \bibfield  {author} {\bibinfo {author} {\bibfnamefont {K.}~\bibnamefont
  {Everschor-Sitte}}, \bibinfo {author} {\bibfnamefont {M.}~\bibnamefont
  {Sitte}}, \bibinfo {author} {\bibfnamefont {T.}~\bibnamefont {Valet}},
  \bibinfo {author} {\bibfnamefont {A.}~\bibnamefont {Abanov}}, \ and\ \bibinfo
  {author} {\bibfnamefont {J.}~\bibnamefont {Sinova}},\ }\href@noop {}
  {\bibfield  {journal} {\bibinfo  {journal} {New Journal of Physics}\ }\textbf
  {\bibinfo {volume} {19}},\ \bibinfo {pages} {092001} (\bibinfo {year}
  {2017})}\BibitemShut {NoStop}%
\end{thebibliography}
\end{document}